\documentclass[11pt,oneside]{article}
\usepackage{jheppub} 
\usepackage[table,xcdraw]{xcolor}
\usepackage{graphicx,latexsym,keyval,ifthen,moreverb,epstopdf}
\usepackage{siunitx}
\usepackage{color}
\usepackage{amsmath}
\usepackage{slashed}
\usepackage{multirow}
\usepackage[T1]{fontenc}
\usepackage{calligra}
\usepackage{subfigure}
\usepackage{pdflscape}
\usepackage{longtable}
\usepackage{cancel}
\usepackage{hyperref}
\usepackage{enumerate}
\usepackage{amsfonts,mathrsfs}

\usepackage{braket}
\usepackage{svg}
\usepackage{tikz}
\usepackage{setspace}
\usepackage{ulem}

\newcommand{\nn}{\nonumber \\}

\newcommand{\beq}{\begin{equation}} 
\newcommand{\eeq}{\end{equation}} 
\newcommand{\ba}{\begin{array}}  
\newcommand{\ea}{\end{array}} 
\newcommand{\bea}{\begin{eqnarray}}  
\newcommand{\eea}{\end{eqnarray} }  
\newcommand{\be}{\begin{eqnarray}}  
\newcommand{\ee}{\end{eqnarray} }  
\newcommand{\bal}{\begin{align}}
\newcommand{\eal}{\end{align}}   
\newcommand{\bi}{\begin{itemize}}  
\newcommand{\ei}{\end{itemize}}  
\newcommand{\ben}{\begin{enumerate}}  
\newcommand{\een}{\end{enumerate}}  
\newcommand{\bc}{\begin{center}}
\newcommand{\ec}{\end{center}} 
\newcommand{\bt}{\begin{table}}
\newcommand{\et}{\end{table}}  
\newcommand{\btb}{\begin{tabular}}
\newcommand{\etb}{\end{tabular}}

\newcommand{\cO}{{\mathcal O}}

\title{EFT analysis of New Physics at COHERENT with Dirac neutrinos}
\author[a]{Víctor Bresó-Pla,}
\author[b]{Sergio Cruz-Alzaga,}
\author[b]{Martín González-Alonso,}
\author[b]{and Suraj Prakash}

\affiliation[a]{Institute for Theoretical Physics, Universit\"at Heidelberg, Germany}
\affiliation[b]{Departament de Física Teòrica, IFIC (Universitat de València - CSIC), Parc Científic UV,\\ C/ Catedrático José Beltrán 2, E-46980 Paterna (Valencia), Spain}

\emailAdd{breso@thphys.uni-heidelberg.de}
\emailAdd{sergio.delacruz.alzaga@ific.uv.es}
\emailAdd{martin.gonzalez@ific.uv.es}
\emailAdd{suraj.prakash@ific.uv.es}

\abstract{We study the sensitivity of COHERENT-like experiments to non-standard contributions within the so-called $\nu$WEFT framework. The latter is the most general low-energy effective field theory that includes not only the light SM fields but also additional right-handed Dirac neutrinos.  Our analysis includes for the first time flavor-general New Physics effects in neutrino production (pion and muon decays) and neutrino detection (through Coherent Elastic Neutrino-Nucleus Scattering). Despite the generality, the results can be written in compact form and are easy to implement in existing or future analyses using effective nuclear charges. We use current COHERENT data to set constraints on the corresponding effective operators, and we estimate the sensitivity of future measurements.}

\begin{document}
\setstretch{1.15}
\maketitle

\section{Introduction}\label{sec:intro}
Precision measurements with neutrinos are important ingredients in the so-called Intensity Frontier, providing stringent tests of the electroweak sector of the Standard Model (SM). By analyzing neutrino interactions with high accuracy, it is possible to probe fundamental parameters such as the weak mixing angle and neutrino couplings to the $Z$ boson. These measurements not only allow for consistency tests within the SM but also offer sensitivity to physics beyond the Standard Model (BSM), where deviations from theoretical predictions may hint at new interactions or particles. The increasing precision of neutrino experiments thus represents a powerful avenue in the search for new phenomena in particle physics.

A particularly recent and exciting example of such precision measurements is Coherent Elastic Neutrino-Nucleus Scattering (CE$\nu$NS). Theoretically predicted in the 1970s~\cite{Freedman:1973yd}, CE$\nu$NS was first observed in 2017 by the COHERENT experiment~\cite{COHERENT:2017ipa} using the Spallation Neutron Source (SNS) at the Oak Ridge National Laboratory. This process occurs when neutrinos scatter off nuclei with a small momentum transfer, leading to a cross section that scales quadratically with the number of nucleons. This coherent enhancement makes CE$\nu$NS a powerful new window into weak interactions, neutrino properties, and potential deviations from the Standard Model. As a result, CE$\nu$NS has become a key target for current and future experimental efforts, aiming for higher precision, different targets, and varied neutrino sources ~\cite{COHERENT:2020iec,COHERENT:2020ybo,COHERENT:2021xmm,Colaresi:2022obx,Ackermann:2025obx,CCM:2021leg,Zema:2020qen}. 
Notably, strong indications suggest that reactor and solar neutrinos have also been recently detected through CE$\nu$NS~\cite{CONNIE:2024pwt,CONUS:2020skt,XENON:2024ijk,PandaX:2024muv}. 

The most precise CE$\nu$NS measurements currently available by far are those obtained by the COHERENT collaboration using Liquid Argon (LAr) and Cesium Iodide (CsI) targets~\cite{COHERENT:2020ybo,COHERENT:2021xmm}. A wide range of theoretical investigations have explored the impact of these measurements on different facets of detection physics~\cite{Barranco:2005yy,Scholberg:2005qs,Cadeddu:2017etk,Papoulias:2017qdn,Shoemaker:2017lzs,Liao:2017uzy,Cadeddu:2018dux,AristizabalSierra:2018eqm,Denton:2018xmq,Altmannshofer:2018xyo,Giunti:2019xpr,Coloma:2019mbs,Skiba:2020msb,Hoferichter:2020osn,Miranda:2020tif,AtzoriCorona:2022qrf,DeRomeri:2022twg,Breso-Pla:2023tnz}.
Notable topics include studying nuclear structure, like the extraction of the so-called neutron skin, probing the electroweak interaction through the determination of the weak mixing angle, constraining electromagnetic neutrino properties, like its magnetic moment, searching for light mediators, and constraining non-standard neutral-current interactions.

In Ref.~\cite{Breso-Pla:2023tnz} we studied the COHERENT measurements, including for the first time nonstandard effects in production and detection, using an Effective Field Theory (EFT) framework with the field content of the SM (that is, only including three left-handed neutrino fields). At low- and high-energies this corresponds to 
the so-called weak EFT (WEFT) and Standard Model EFT (SMEFT) respectively. In that scenario we found that the BSM production effects either cancel or are highly suppressed, supporting the usual approach of neglecting them. On the other hand, the sensitivity to BSM effects in detection is not suppressed and was found to be an important input for SMEFT global fits of Electroweak Precision Observables.

In a recent letter~\cite{Breso-Pla:2025cul}, we demonstrated that the inclusion of right-handed neutrinos dramatically changes the picture: COHERENT does exhibit an interesting sensitivity to nonstandard effects in the production involving pion and muon decays. To illustrate the main point in simple terms, Ref.~\cite{Breso-Pla:2025cul} focuses on the study of the interactions mediating muon decay assuming lepton flavor conservation. This article presents the results obtained in the most general low-energy EFT that includes right-handed neutrinos, lepton-flavor violation (LFV), and non-standard effects in muon decay, pion decay, and CE$\nu$NS detection.  

The structure of this article is as follows. Sec.~\ref{sec:eft} presents the EFT operators that contribute to the relevant (anti)neutrino detection and production processes. 
Sec.~\ref{sec:amplitudes} discusses the associated amplitudes and phase space integration. Using them we calculate the observable of interest --i.e., the COHERENT event rate-- in Sec.~\ref{sec:rates} and outline interesting limits of our general result. In Sec.~\ref{sec:pheno} we present results from a statistical analysis of COHERENT data to extract the values of relevant theoretical parameters. Finally, we offer our conclusions in Sec.~\ref{sec:conclusions}.

\section{EFT framework}\label{sec:eft}
\subsection{Quark-level EFT}\label{subsec:quark-lepton-eft}
We consider the weak effective field theory extended with $3$ families\footnote{Our final results will be written in matricial form and they will be valid for any number of $\nu_R$ fields. Intermediate results assume 3 $\nu_R$ fields, but they are trivially generalizable to $n$ fields (one only needs to use the $\nu_R$ mass basis so that the $U$ matrix is absent).} of right-handed neutrino, called $\nu$WEFT (also known as $\nu$LEFT and LNEFT)~\cite{Chala:2020vqp,Li:2020lba}.  
We present below the lepton-number conserving interactions that are relevant for neutrino production and detection processes at COHERENT. Equivalently, we are working with the SMNEFT at dimension 6~\cite{Chala:2020vqp,Li:2020lba}, since lepton-number-violating operators are not generated at that order.

At the detection level the relevant interactions are QED, along with dim-5 and dim-6 $\nu$WEFT terms
\begin{eqnarray}\label{eq.Lag-detection}
    \mathcal{L}
    &\supset& {\cal L}_{\rm QED} -\frac{\left[\mu_{\nu}\right]_{\alpha\beta}}{2}(\bar{\nu}_\alpha \sigma^{\mu\nu}P_R \nu_\beta)F_{\mu\nu}
    + h.c. -\,\sum\limits_{q=u,d}\,\frac{\mu_{q}}{2}(\bar{q}\sigma^{\mu\nu}P_R q) F_{\mu\nu} + h.c. \nn
    &&- \frac{1}{v^2}\sum_{q=u,d}\left\{\left[g^{\,qq}_V\mathbb{I}+\epsilon^{\,qq}_{V}\right]_{\alpha\beta}(\bar{q}\gamma^\mu q)(\bar{\nu}_\alpha \gamma_\mu P_L \nu_\beta) +\left[\widetilde{\epsilon}^{\,qq}_{V}\right]_{\alpha\beta}(\bar{q}\gamma^\mu q)(\bar{\nu}_\alpha \gamma_\mu P_R \nu_\beta) \right. \nn
    &&\hspace{1.6cm}   +\left[g^{\,qq}_A\mathbb{I}+\epsilon^{\,qq}_{A}\right]_{\alpha\beta}(\bar{q}\gamma^\mu\gamma^5 q)(\bar{\nu}_\alpha \gamma_\mu P_L \nu_\beta) + \left[\widetilde{\epsilon}^{\,qq}_{A}\right]_{\alpha\beta}(\bar{q}\gamma^\mu\gamma^5 q)(\bar{\nu}_\alpha \gamma_\mu P_R \nu_\beta) \nn
    &&\hspace{1.6cm}   +\left[\widetilde{\epsilon}^{\,qq}_{S}\right]_{\alpha\beta}(\bar{q} q)(\bar{\nu}_\alpha P_R \nu_\beta) + h.c. + \left[\widetilde{\epsilon}^{\,qq}_{P}\right]_{\alpha\beta}(\bar{q}i\gamma^5 q)(\bar{\nu}_\alpha  P_R \nu_\beta)+ h.c. \nn
    &&\hspace{1.6cm} \left. + \left[\widetilde{\epsilon}^{\,qq}_{T}\right]_{\alpha\beta}(\bar{q}\sigma^{\mu\nu} q)(\bar{\nu}_\alpha \sigma_{\mu\nu}P_R \nu_\beta)+ h.c. \right\} ~
\end{eqnarray}
Here, $v \equiv (\sqrt{2} G_F)^{-1/2}\approx 246.22$~GeV, where $G_F\approx 1.166\times 10^{-5}~{\rm GeV}^{-2}$~\cite{ParticleDataGroup:2024cfk} is the measured value of the Fermi constant. The tree-level values of the SM couplings are 
$g_V^{qq} = T^3_q - 2 Q_q \sin^2 \theta_W$ and $g_A^{qq}   = - T^3_q$, 
where $\theta_W$ is the weak mixing angle and $T_q^3$ and $Q_q$ are the weak isospin and electromagnetic charges of quark $q$. We use the tilde to mark the Wilson Coefficients (WCs) of 4-fermion operators that involve $\nu_R$, which are thus new with respect to Ref.~\cite{Breso-Pla:2023tnz}. For the neutrino magnetic moment we follow the usual notation~\cite{Giunti:2024gec}.

We now move on  to (anti)neutrino production. The $\nu$WEFT interactions relevant for pion decay are
\begin{eqnarray}\label{eq.Lag-pion-decay}
        \mathcal{L}
        \supset-\frac{2 V_{ud}}{v^2}&&\left\{\left[\mathbb{I}+\epsilon^{ud}_{L}\right]_{\alpha\beta}(\bar{u}\gamma^\mu P_L d)(\bar{l}_\alpha \gamma_\mu P_L \nu_\beta) +\left[\epsilon^{ud}_{R}\right]_{\alpha\beta}(\bar{u}\gamma^\mu P_R d)(\bar{l}_\alpha \gamma_\mu P_L \nu_\beta) \right.\nn
        &&+\left[\widetilde{\epsilon}^{\,ud}_{L}\right]_{\alpha\beta}(\bar{u}\gamma^\mu P_L d)(\bar{l}_\alpha \gamma_\mu P_R \nu_\beta) +\left[\widetilde{\epsilon}^{\,ud}_{R}\right]_{\alpha\beta}(\bar{u}\gamma^\mu P_R d)(\bar{l}_\alpha \gamma_\mu P_R \nu_\beta)\nn
        &&+\frac{1}{2}\left[\epsilon^{ud}_{S}\right]_{\alpha\beta}(\bar{u} d)(\bar{l}_\alpha P_L \nu_\beta)  - \frac{1}{2}\left[\epsilon^{ud}_{P}\right]_{\alpha\beta}(\bar{u}\gamma^5 d)(\bar{l}_\alpha  P_L \nu_\beta)\nn
        &&+\frac{1}{2}\left[\widetilde{\epsilon}^{\,ud}_{S}\right]_{\alpha\beta}(\bar{u} d)(\bar{l}_\alpha P_R \nu_\beta)  - \frac{1}{2}\left[\widetilde{\epsilon}^{\,ud}_{P}\right]_{\alpha\beta}(\bar{u}\gamma^5 d)(\bar{l}_\alpha  P_R \nu_\beta)\nn
        &&+\left.\frac{1}{4}\left[\epsilon^{\,ud}_{T}\right]_{\alpha\beta}(\bar{u}\sigma^{\mu\nu}P_L d)(\bar{l}_\alpha \sigma_{\mu\nu}P_L \nu_\beta)+\frac{1}{4}\left[\widetilde{\epsilon}^{\,ud}_{T}\right]_{\alpha\beta}(\bar{u}\sigma^{\mu\nu}P_R d)(\bar{l}_\alpha \sigma_{\mu\nu}P_R \nu_\beta)\right\} \nn 
        && + ~h.c.,
\end{eqnarray}
where $V_{ud}$ is the (1,1) element of the Cabibbo--Kobayashi--Maskawa~(CKM) matrix, and $\alpha,\beta$ are lepton flavor indices.

The $\nu$WEFT interactions relevant for muon decay are 
\begin{align}\label{eq:Lag-muon-decay}
    \mathcal{L}
    &\supset 
    -\frac{2}{v_0^2}\,\sum_{X,\eta,\epsilon}\,
    [\hat{h}^{X}_{\epsilon \eta}]_{\alpha \beta}(\bar{e}_{\epsilon}\Gamma^X (\nu_{\beta})_{\rho})((\bar{\nu}_{\alpha})_{\gamma} \Gamma_X \mu_\eta)\,+\,h.c.\nn
    &=-\frac{2}{v^2}\,\sum_{X,\eta,\epsilon}\,
    [h^{X}_{\epsilon \eta}]_{\alpha \beta}(\bar{e}_{\epsilon}\Gamma^X (\nu_{\beta})_{\rho})((\bar{\nu}_{\alpha})_{\gamma} \Gamma_X \mu_\eta)\,+\,h.c.~,
\end{align}
where $v_0$ is the tree-level value of the Higgs vacuum expectation value, $X$ runs over the different Lorentz structures ($\Gamma_X=\{\mathbb{I},\,\gamma_\mu,\,\frac{1}{\sqrt{2}}\sigma_{\mu\nu}\}$ for $X=\{S,V,T\}$ respectively), $\alpha,\beta$ are the flavors of the (anti)neutrinos, and $\eta,\epsilon,\rho,\gamma$ are the field chiralities. Note that for a given Lorentz structure, only two of the chiralities are independent, which is why the sum runs only over $\eta$ and $\epsilon$. 
Finally, two of the tensor couplings are identically zero due to the Fierz identities: $h_{LL}^{T}\equiv h_{RR}^{T}\equiv 0$ (likewise for the hatted variables). 

It is customary to use $v$ (instead of $v_0$) as the normalizing scale in the muon-decay Lagrangian~\cite{Fetscher:1986uj,Fetscher:1994nv,Pich:2013lsa,MuonDecayParameters:2024cfk}, as we have done in the second line of Eq.~(\ref{eq:Lag-muon-decay}). Since $v$ is defined via $G_F$, which is extracted from the muon decay rate itself, the associated Wilson Coefficients $h^{X}_{\epsilon \eta}$ are not independent, but satisfy the following normalization condition\footnote{Indeed, the $h$ and $\hat h$ variables are related by $h^X_{\epsilon\eta}=\hat h^X_{\epsilon\eta}/\sum_{\epsilon,\eta}{\rm Tr}\left[
    \frac{1}{4}\, \hat h^{S}_{\epsilon\eta}(\hat h^{S}_{\epsilon\eta})^\dagger 
    \,+\, \hat h^{V}_{\epsilon\eta}(\hat h^{V}_{\epsilon\eta})^\dagger
    \,+\,3 \,\hat h^{T}_{\epsilon\eta}(\hat h^{T}_{\epsilon\eta})^\dagger \right]$.
}
\begin{align}
\label{eq:muon_decay_norm}
    \sum_{\epsilon,\eta}{\rm Tr}\left[
    \frac{1}{4}\, h^{S}_{\epsilon\eta}(h^{S}_{\epsilon\eta})^\dagger 
    \,+\, h^{V}_{\epsilon\eta}(h^{V}_{\epsilon\eta})^\dagger
    \,+\,3 \,h^{T}_{\epsilon\eta}(h^{T}_{\epsilon\eta})^\dagger \right] = 1~,
\end{align}
which in particular implies that 
$|[h^{V}_{\epsilon \eta}]_{\alpha \beta}|\leq 1$, $|[h^{S}_{\epsilon \eta}]_{\alpha \beta}|\leq 2$ and $|[h^{T}_{\epsilon \eta}]_{\alpha \beta}|\leq 1/ \sqrt{3}$. The connection with the $g^X_{\eta\epsilon}$ couplings used in the Particle Data Group (PDG) review on muon-decay parameters~\cite{Fetscher:1986uj,MuonDecayParameters:2024cfk}, and in our recent letter~\cite{Breso-Pla:2025cul}, is simply 
$[h^{X}_{\epsilon\eta}]_{\alpha\beta} = g^X_{\epsilon\eta} \,\delta_{\alpha\mu}\,\delta_{\beta e}$, i.e., lepton-flavor conservation is assumed in that approach. 
The relation with the coefficients in Ref.~\cite{Breso-Pla:2023tnz} is $[\hat{h}^{V}_{LL}]_{\alpha \beta} = \delta_{\mu\alpha } \delta_{\beta e} + [\rho_{L}]^*_{\mu \alpha \beta e}$, and 
$[\hat{h}^{S}_{RR}]_{\alpha \beta} = -2\, [\rho_R]^*_{\mu \alpha \beta e}$,
whereas the operators with right-handed neutrinos were not present in that work. 
Finally, the SM limit is recovered when all coefficients vanish except $[\hat{h}^{V}_{LL}]_{\mu e} =[h^{V}_{LL}]_{\mu e} =1$.
    
In the subsets of the $\nu$WEFT Lagrangian, presented in Eqs.~\eqref{eq.Lag-detection}-\eqref{eq:Lag-muon-decay}, the charged fields (the $u$, $d$ quarks and the charged leptons $\ell_{\alpha}$) are in a basis where their kinetic and mass terms are diagonal, while an additional rotation is needed to make the neutrino mass matrix diagonal: $\nu_{\alpha} = \sum_{n=1}^3 U_{\alpha n} \nu_n$. In other words, the $\nu_L$ fields are in the flavor basis. On the other hand, the $\nu_R$ fields can be chosen to be in the mass basis, $\nu_{R,n}$ (i.e., no $\nu_R$ rotation is needed to make the neutrino mass matrix diagonal), and then the $\nu_L$ rotation matrix $U$ is simply the Pontecorvo-Maki-Nakagawa-Sakata (PMNS) matrix, as often done in the quark sector at the SMEFT level with the CKM matrix. 
In the case of 3 $\nu_R$ states, one can also write the $\nu$WEFT Lagrangian in terms of $\nu_R$ flavor states, defined as $\nu_{R,\alpha}\equiv U_{\alpha n}\nu_{R,n}$. We will work with this basis, which has the benefits of yielding very similar expressions for $\nu_L$ and $\nu_R$ amplitudes and working in the flavor basis for both $\nu_L$ and $\nu_R$. It must be emphasized that this choice of basis is irrelevant for our final results since the $U$ factors cancel during the calculation of the rate.

In full generality, $\epsilon^{\,qq}_{V,A}$ and $\widetilde\epsilon^{\,qq}_{V,A}$ are Hermitian matrices in the neutrino indices, whereas the remaining Wilson Coefficients are general complex matrices.

\subsection{Nucleon-level EFT}\label{subsec:nucleon-lepton-eft}
For the description of CE$\nu$NS interactions, introducing an effective Lagrangian with nucleons, rather than quarks, serves as a convenient  
intermediate step. The neutral current (NC) interactions relevant for our study in the relativistic nucleon-level EFT are
\begin{eqnarray}\label{eq:lag-nucleon-lepton}
\mathcal{L}_{N\nu}
\label{eq:L_Nnu}
& \supset & 
\,e\,(\bar{p}\gamma^\mu p)A_\mu - \frac{\left[\mu_{\nu}\right]_{\alpha\beta}}{2}(\bar{\nu}_\alpha \sigma^{\mu\nu}P_R \nu_\beta)F_{\mu\nu} + h.c. -\,\sum\limits_{N=p,n}\,\frac{\mu_{N}}{2}(\bar{N}\sigma^{\mu\nu}P_R N) F_{\mu\nu} + h.c. \nn
&&
-\frac{1}{2v^2} \sum\limits_{N=p,n} \left\{   
\left[g^{\,\nu N}_V\right]_{\alpha\beta}(\bar{N}\gamma^\mu N)(\bar{\nu}_\alpha \gamma_\mu P_L \nu_\beta) +\left[\widetilde{g}^{\,\nu N}_{V}\right]_{\alpha\beta}(\bar{N}\gamma^\mu N)
(\bar{\nu}_\alpha \gamma_\mu P_R \nu_\beta)\right. \nn 
&&\hspace{1.65cm} +\left[g^{\,\nu N}_A\right]_{\alpha\beta}(\bar{N}\gamma^\mu\gamma^5 N)(\bar{\nu}_\alpha \gamma_\mu P_L \nu_\beta) +\left[\widetilde{g}^{\,\nu N}_{A}\right]_{\alpha\beta}(\bar{N}\gamma^\mu\gamma^5 N)(\bar{\nu}_\alpha \gamma_\mu P_R \nu_\beta)\nn
&&\hspace{1.65cm}+\left[\widetilde{g}^{\,\nu N}_{S}\right]_{\alpha\beta}(\bar{N} N)(\bar{\nu}_\alpha P_R \nu_\beta) + h.c. + \left[\widetilde{g}^{\,\nu N}_{P}\right]_{\alpha\beta}(\bar{N}i\gamma^5 N)(\bar{\nu}_\alpha  P_R \nu_\beta)+ h.c.\nn
&&\hspace{1.65cm}\left.+\left[\widetilde{g}^{\,\nu N}_{T}\right]_{\alpha\beta}(\bar{N}\sigma^{\mu\nu} N)(\bar{\nu}_\alpha \sigma_{\mu\nu}P_R \nu_\beta)+ h.c. \right\}~.
\end{eqnarray}
We include the nucleon-photon interaction ($\mu_N$ denotes the magnetic moment of the nucleon $N$), since it is relevant for photon-mediated neutrino-nucleon scattering and it also involves the neutrino magnetic moment operator. 

The relevant neutrino-nucleon couplings are given, in terms of the $\nu$WEFT WCs, as
\begin{eqnarray}
g_V^{\,\nu p} &=& 2\left[(2g_V^{uu}+g_V^{dd})\mathbb{I}+(2\epsilon_V^{uu}+\epsilon_V^{dd})\right], \qquad \widetilde{g}_V^{\,\nu p} = 2\left[2\widetilde{\epsilon}_V^{\,uu}+\widetilde{\epsilon}_V^{\,dd}\right], \nn 
g_V^{\,\nu n} &=& 2\left[(g_V^{uu}+2g_V^{dd})\mathbb{I}+(\epsilon_V^{uu}+2\epsilon_V^{dd})\right], \qquad
\widetilde{g}_V^{\,\nu n} = 2\left[\widetilde{\epsilon}_V^{\,uu}+2\widetilde{\epsilon}_V^{\,dd}\right],\nn
\widetilde{g}_S^{\,\nu p} &=& 2m_p\left[\frac{f_u^p}{m_u}\widetilde{\epsilon}_S^{\,uu}+\frac{f_d^p}{m_d}\widetilde{\epsilon}_S^{\,dd}\right], \nn
\widetilde{g}_S^{\,\nu n} &=& 2m_n\left[\frac{f_u^n}{m_u}\widetilde{\epsilon}_S^{\,uu}+\frac{f_d^n}{m_d}\widetilde{\epsilon}_S^{\,dd}\right]~,
\end{eqnarray}
The couplings $g_A^{\,\nu N}$ and $\widetilde{g}_X^{\,\nu N} (X=A,\,P, \,T)$ are defined analogously but, as we shall show imminently, these operators are not relevant for the remainder of our calculations. 
Finally, $f^N_q\equiv f^N_q(0)$, where the corresponding form factor is defined by\footnote{The $q^2\equiv(p-p')^2$ dependence of the form factor is captured by higher dimension operators in the nucleon-level EFT. At $q^2=0$ one has $\{ f^p_u,~f^p_d,~f^n_u,~f^n_d\}=\{20.8\pm 1.5,~ 41.1\pm 2.8,~ 18.9\pm 1.4,~ 45.1\pm 2.7\}\times 10^{-3}$~\cite{Hoferichter:2015dsa}.}
\begin{equation}
    \langle N(p') |m_q\bar{q}q|N(p)\rangle = m_N f^N_q (q^2) \bar{N}(p')N(p)~.
\end{equation}

Since the 3-momentum transfer in CE$\nu$NS processes is much smaller than the nucleon mass, the nucleons can be described by non-relativistic quantum fields $\psi_N$, $N=p,n$ (2-component spinors) whereas relativistic fields should still be used for the light degrees of freedom (neutrino and photon). The corresponding EFT Lagrangian is thus organized as an expansion in $\boldsymbol{\nabla}/m_N$, where $\boldsymbol{\nabla}$ denotes spatial derivatives. This setup, known as the pionless EFT~\cite{vanKolck:1999mw}, is helpful to relate the nucleon-level EFT to the observables, as this enables the calculation for nuclei with arbitrary spin. We will use it in this work to describe CE$\nu$NS detection at leading order in the non-relativistic expansion. At this order, the nucleon bilinears take the following form
\begin{eqnarray}
\bar{N}N&\longrightarrow&\psi_N^\dagger\psi_N~, \hspace{2.05cm} \bar{N}\gamma^0N \longrightarrow \psi_N^\dagger\psi_N~,\nn \bar{N}\gamma^k\gamma^5N&\longrightarrow&\psi_N^\dagger\sigma^k\psi_N~, \qquad\qquad
\bar{N}\sigma^{kl}N \longrightarrow \epsilon^{klm}\psi_N^\dagger\sigma^m\psi_N~,
\end{eqnarray}
whereas the remaining bilinears vanish. Here $\sigma^k$ are the Pauli matrices and $\epsilon^{klm}$ the antisymmetric Levi-Civita tensor. 
This shows that the scalar (tensor) operator gives the same structure as the vector (axial) one\footnote{It is well-known that the same happens in $\beta$ decay, where scalar and vector (tensor and axial) operators give the same structure, which leads to the Fermi (Gamow-Teller) matrix element~\cite{Bethe:1936zz}. See Ref.~\cite{Falkowski:2021vdg} for a recent study of subleading recoil effects using the same non-relativistic EFT approach as employed here.}, and thus they are (not) coherently enhanced, contrary to the results in the first CE$\nu$NS studies of tensor interactions~\cite{Barranco:2011wx,Healey:2013vka,Papoulias:2015iga,Papoulias:2017qdn,AristizabalSierra:2018eqm}. The absence of coherent enhancement for the tensor interaction, relevant also in the context of dark matter direct detection and $\mu\to e$ conversion, was correctly addressed in Refs.~\cite{Altmannshofer:2018xyo,Hoferichter:2020osn,Glick-Magid:2023uhk}\footnote{Recoil corrections generated by tensor interactions are coherently enhanced~\cite{Liao:2025hcs} but are also much smaller than current experimental uncertainties.} using theoretical frameworks different to ours. 
However, these works seemed unaware that this issue had been previously misaddressed in the CE$\nu$NS literature, and the point did not appear to have been conveyed for some time, resulting in the persistence of the error~\cite{AristizabalSierra:2019zmy,Bischer:2019ttk,Papoulias:2019xaw,Chang:2020jwl,Han:2020pff,Li:2020lba,CONUS:2021dwh,Majumdar:2021vdw,AristizabalSierra:2022axl,Chen:2022xkk,Majumdar:2022nby,DeRomeri:2022twg,Fridell:2023rtr,Demirci:2023tui,Chatterjee:2024vkd,Karmakar:2024ywn} until very recently~\cite{DeRomeri:2024iaw,Chattaraj:2025rtj}\footnote{The preliminary results of this work, including the absence of coherent enhancement for the tensor interaction, were presented in the 2024 Magnificent CE$\nu$NS workshop~\cite{Sergio_talk_2024}.}.
Finally, the lack of coherent enhancement of the tensor bilinear also suppresses the contribution of the $\mu_N-\mu_\nu$ interaction via a mediating photon.

All in all, the non-relativistic Lagrangian is
\begin{eqnarray}
    \label{eq:L_NR}
    {\cal L}_{\rm NR} &\supset& \, e A_0\,(\psi_p^\dagger \psi_p) - \frac{1}{2v^2} \sum_{N = p,n} (\psi_N^\dagger \psi_N)\Big\{\left[g^{\,\nu N}_V\right]_{\alpha\beta}(\bar{\nu}_\alpha \gamma_0 P_L \nu_\beta) \nn
    && +\left[\widetilde{g}^{\,\nu N}_{V}\right]_{\alpha\beta}
    (\bar{\nu}_\alpha \gamma_0 P_R \nu_\beta) + \left[\widetilde{g}^{\,\nu N}_{S}\right]_{\alpha\beta}(\bar{\nu}_\alpha P_R \nu_\beta) + h.c. \Big\} + \cO(\nabla/m_N),
    \end{eqnarray}
plus the neutrino magnetic moment term in Eq.~\eqref{eq.Lag-detection}, which is unchanged by the non-relativistic reduction since it does not involve nucleons. The term containing $A_0$ is obtained from the non-relativistic reduction of the proton-photon coupling  shown in Eq.~\eqref{eq:L_Nnu}. 
We also omit the nucleon magnetic moment terms, since they have tensorial structure. Finally, we do not consider in this work subleading interactions with more than two nucleon fields, e.g., two-body currents.
    
Let us now discuss the nuclear matrix elements of non-relativistic nucleon currents ($\psi_N^\dagger \psi_N$). 
For an initial (final) nucleus ${\cal N}$ with charge $Z$, mass number $A$, momentum $k$ ($k'$),  spin $J$, and  spin projection along the z-axis $J_z$ ($J'_z$), in the low-recoil limit, the matrix elements (invariant with respect to rotational and isospin symmetry) take the form~\cite{Breso-Pla:2023tnz}
\begin{eqnarray}\label{eq:nuclear-matrix-elem}
\bra{ {\cal N}(k^\prime, J_z^\prime) } \psi_p^\dagger \psi_p  \ket{ {\cal N}(k, J_z) }  &\approx&  2 m_{\cal N} \,  \mathcal{F}_{p} (q^2)  Z \,\delta_{J_z^\prime J_z} , \nn
\bra{ {\cal N}(k^\prime, J_z^\prime) } \psi_n^\dagger \psi_n  \ket{ {\cal N}(k, J_z) }  &\approx&  2 m_{\cal N} \,  \mathcal{F}_{n} (q^2)  (A-Z)\,\delta_{J_z^\prime J_z}~,
\end{eqnarray} 
where we find the famous coherent enhancement, and the nuclear form factors $\mathcal{F}_N (q^2)$, where $q^2\equiv (k -  k^\prime)^2$. From isospin symmetry it follows that $\mathcal{F}_N (0)=1$, and we will approximate ${\cal F}_p(q^2) = {\cal F}_n(q^2) \equiv {\cal F}(q^2)$.

\section{Amplitudes}\label{sec:amplitudes}
\subsection{From the amplitudes to the rate}\label{subsec:amps-to-rates}

We follow the approach presented in Ref.~\cite{Breso-Pla:2023tnz}  to calculate the event rate for neutrinos emitted through $\pi^+\to \mu^+\nu_{k}$ or $\mu^+\to e^+\bar{\nu}_m\nu_k$ and detected via $\nu_{k}{\cal N}\to \nu_{j}{\cal N}$, where ${\cal N}$ is the target nucleus. The event rate per unit time $t$, incident neutrino energy $E_{\nu}$, nuclear recoil energy $T$ and target particle, 
is related to the production and detection QFT amplitudes, denoted by ${\cal M}_{\alpha k}^P\equiv {\cal M}\left( S \to X_{\alpha} \nu_{k} \right)$ and ${\cal M}_{jk}^D\equiv {\cal M}\left(  \nu_{k} {\cal N} \to \nu_{j} {\cal N}  \right)$, by~\cite{Breso-Pla:2023tnz}\footnote{The event rate for antineutrinos is defined analogously.}
\begin{equation}
\label{eq:DifferentialRateEvents}
    \frac{1}{N_T}\frac{dN^S_\alpha}{dt \, dE_{\nu}\, dT}
    = \frac{N_S(t)}{32 \pi L^{2} m_S m_{\cal N} E_{\nu}} 
    \sum_{j,k,l}
    e^{-i\frac{L \Delta m_{kl}^{2}}{2E_{\nu}}} \int d \Pi_{P^{\prime}} \mathcal{M}^{P}_{\alpha k} \bar{\mathcal{M}}^{P}_{\alpha l} \int d \Pi_{D^{\prime}} \mathcal{M}^{D}_{j k} \bar{\mathcal{M}}^{D}_{j l} ~,
\end{equation}
where $S=\pi^+,\mu^+$ is the source particle, $N_T$ is the number of target particles, $m_{S}(m_{\cal N})$ parametrize the mass of the source (target) particle, the bar over the amplitudes indicates complex conjugation, and $\Delta m_{kl}^{2}\equiv m_{k}^{2}-m_{l}^{2}$ is the mass squared difference between neutrino (mass) eigenstates. The latter appears in the formula through the familiar $e^{-i \,L \Delta m_{kl}^2 / (2E_\nu)}$ oscillatory factor, which can be approximated as one due to the short baseline of COHERENT. The phase space elements, {$d \Pi \equiv \frac{d^3 k_1}{(2 \pi)^3 2 E_1} \dots  {d^3 k_n \over (2 \pi)^3 2 E_n} (2\pi)^4 \delta^4(\sum p_n - \sum k_i )$}, are defined as usual (in terms of the 4-momenta $k_i$ and energies $E_i$ of the final states and the total 4-momentum of the initial state $\sum p_n$) for production ($d \Pi_{P}$) and detection processes ($d \Pi_{D}$). The primed phase space elements are defined by $d \Pi_{P}\equiv d \Pi_{P^{\prime}} dE_{\nu} $ and $d \Pi_{D}\equiv d \Pi_{D^{\prime}} dT $. The integral sign describes integration as well as sums and averages over spin (and other unobserved  degrees of freedom). 
Finally, the number of source particles $S$ is described by the time-dependent function $N_S (t)=n_{\rm POT}\,r_{S/p}\,\tau_S \,g_S(t)$. Here, $n_{\rm POT}$ is the total number of protons on target delivered at the SNS, $r_{S/p}$ is the number of $S$ particles produced per proton, $\tau_S$ is the $S$ lifetime and $g_S(t)$ encodes the $N_S$ time dependence.  

Compared to Ref.~\cite{Breso-Pla:2023tnz}, our field content now also includes right-handed neutrinos. Since the sum over $j,k$ indices runs over helicity states, we have to consider now not only the cases where $j,k=1,2,3$ for left-handed neutrinos, but also for the right-handed ones. We note that in the limit of zero neutrino masses, chirality and helicity coincide and left-right interference terms vanish.

\subsection{Detection}\label{subsec:detection-amps}

\begin{figure}
\centering
\includegraphics[width=0.95\linewidth]{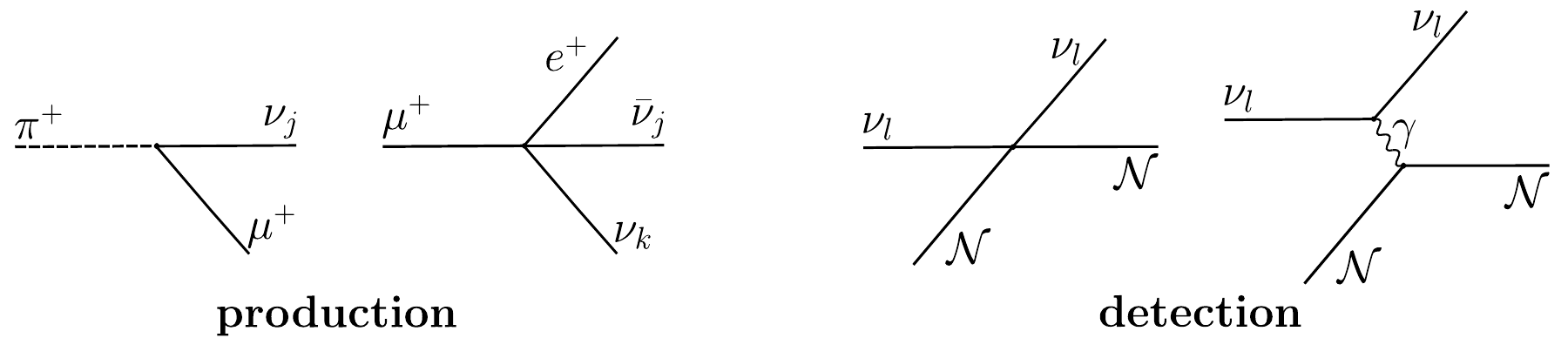}
\caption{Feynman diagram for production and detection (we include the contact interaction and the photon-mediated topologies). Analogous diagrams apply for antineutrino detection.}
\label{fig:feynman-diagrams}
\end{figure}

The detection of (anti)neutrinos via CE$\nu$NS occurs through (i) a four-fermion contact interaction involving (anti)neutrinos and nuclei and (ii) a photon-mediated tree-level process,  {\it cf.} Fig.~\ref{fig:feynman-diagrams}. The corresponding detection amplitude can be written in terms of the nuclear matrix elements outlined in Eq.~\eqref{eq:nuclear-matrix-elem}, bilinears constituted of Dirac spinors for the neutrino fields, and the photon propagator for the contribution from the neutrino magnetic moment. 
The amplitudes with left- and right-handed neutrinos can be expressed as%
\footnote{If one works instead with $\nu_R$ fields defined in the mass basis of the $\nu$WEFT Lagrangian, then the $U$ matrix should be omitted in the terms associated to $\nu_R$ states.}
\begin{align}
    \left.{\cal M}^{D}_{jk}\right\vert_{\nu_L}
    =&-\frac{1}{v^2}  m_{\cal N}{\cal F}(q^2) \,\delta_{J'_z J_z} \left[U^\dagger \bigg\{ {\cal Q}_V (\bar{u}_{\nu_j} \gamma_0P_L u_{\nu_k}) + {\cal Q}_S^\dagger (\bar{u}_{\nu_j} P_L u_{\nu_k})\right.\nn
    &\qquad\qquad\qquad\qquad\qquad\qquad
    \left. +
    v\, {\cal Q}_F^\dagger\frac{-iq_\mu}{q^2}(\bar{u}_{\nu_j}\sigma^{\mu0} P_L u_{\nu_k})\bigg\}U\right]_{jk}~, \nn
    \left.{\cal M}^{D}_{jk}\right\vert_{\nu_R}
    =&-\frac{1}{v^2}  m_{\cal N}{\cal F}(q^2) \,\delta_{J'_z J_z}\left[U^\dagger \bigg\{ {\cal Q}_{\widetilde V} (\bar{u}_{\nu_j} \gamma_0P_R u_{\nu_k}) + {\cal Q}_S (\bar{u}_{\nu_j} P_R u_{\nu_k})\right. \nn
    &\qquad\qquad\qquad\qquad\qquad\qquad + \left.
    v\,{\cal Q}_F \frac{-iq_\mu}{q^2}(\bar{u}_{\nu_j}\sigma^{\mu0} P_R u_{\nu_k})\bigg\}U\right]_{jk}~,
\label{eq:detection_amplitudes}
\end{align}
where $\left.{\cal M}^{D}_{jk}\right\vert_x\equiv {\cal M}(x_k {\cal N}  \to \nu_j {\cal N})$, for $x=\nu_L,\nu_R$, $u_f$ are the Dirac spinors corresponding to the neutrino fields, and we have introduced the following nuclear charges
\begin{eqnarray}\label{eq:nuclear-charges-definition}
         {\cal Q}_V &=& Z\,g_V^{\nu p}+(A-Z)\,g_V^{\,\nu n}, \qquad {\cal Q}_{\widetilde V} = Z\,\widetilde{g}_V^{\,\nu p}+(A-Z)\,\widetilde{g}_V^{\,\nu n},\nn
         {\cal Q}_S &=& Z\,\widetilde{g}_S^{\,\nu p}+(A-Z)\,\widetilde{g}_S^{\,\nu n}, \qquad {\cal Q}_F = 2\,Z e\,v \mu_\nu 
         = \frac{4v\,\pi\alpha\,Z}{m_e} \frac{\mu_\nu}{\mu_B}~.
\end{eqnarray}
In the last expression, we have normalized the neutrino magnetic moment $\mu_\nu$ with the Bohr magneton $\mu_{_B}\equiv e/2m_e$. 
In the SM case all nuclear charges vanish except ${\cal Q}_V = Q_{\text{SM}} \times \mathbb{I}$, where
\begin{align}
    Q_{\text{SM}} &\equiv \mathcal{Z}(1-4\sin^2\theta_W)-(A-\mathcal{Z}) \simeq - (A-\mathcal{Z})~.
\end{align}
The amplitudes for anti-neutrino detection can be obtained from the neutrino ones in Eq.~(\ref{eq:detection_amplitudes}) by trading the $u_f$ spinors for $v_f$ spinors, exchanging the $k$ and $j$ indices and replacing ${\cal Q}_X \leftrightarrow {\cal Q}_X^\dagger$.

The integrated squared amplitudes are given by
\begin{eqnarray}
    \left.\int d\Pi_{D'}\mathcal{M}^D_{jk}\Bar{\mathcal{M}}^D_{jl}\right\vert_{\nu_L} &=& \frac{\mathcal{F}^2(q^2)\,m_{\cal N}^2 E_\nu}{2\pi v^4} \bigg[U^\dagger\left\{ {\cal Q}_V^2 \cdot \left(1-\frac{m_\mathcal{N}T}{2E^2_\nu}-\frac{T}{E_\nu}\right)  \right.\nn
    &&\qquad\qquad +\,\, {\cal Q}_S {\cal Q}_S^\dagger\cdot\frac{m_\mathcal{N} T}{2E^2_\nu} +  {\cal Q}_F {\cal Q}_F^\dagger\cdot\frac{v^2}{2m_\mathcal{N} T}\left(1-\frac{T}{E_\nu}\right) \nn
    &&\qquad\qquad  + \left.\left({\cal Q}_S {\cal Q}_F^\dagger + {\cal Q}_F {\cal Q}_S^\dagger \right) \cdot \frac{v}{2E_\nu}\left(1-\frac{T}{2E_\nu}\right)\right\}U\bigg]_{lk},\nonumber
    \end{eqnarray}
\begin{eqnarray}
    \label{eq.AmpNRH}
    \left.\int d\Pi_{D'}\mathcal{M}^D_{jk}\Bar{\mathcal{M}}^D_{jl}\right\vert_{\nu_R} &=& \frac{\mathcal{F}^2(q^2)\,m_{\cal N}^2 E_\nu}{2\pi v^4} \bigg[U^\dagger \left\{ {\cal Q}_{\widetilde V}^2\cdot\left(1-\frac{m_\mathcal{N}T}{2E^2_\nu}-\frac{T}{E_\nu}\right)\right.\nn
    && \qquad\qquad +\,\, {\cal Q}_S^\dagger {\cal Q}_S \cdot\frac{m_\mathcal{N} T}{2E^2_\nu}+  {\cal Q}_F^\dagger {\cal Q}_F \cdot\frac{v^2}{2m_\mathcal{N} T}\left(1-\frac{T}{E_\nu}\right)\nn
    && \qquad\qquad +\left. \left({\cal Q}_S^\dagger {\cal Q}_F + {\cal Q}_F^\dagger {\cal Q}_S \right)\cdot\frac{v}{2E_\nu}\left(1-\frac{T}{2E_\nu}\right)\right\}U\bigg]_{lk},
\end{eqnarray}
where we introduced the kinematic recoil energy $T=E_{\cal N}' - m_{\cal N} = -q^2/(2m_{\cal N})$. The integrated squared amplitudes for the detection of anti-neutrinos can be obtained from the above by exchanging the $l$ and $k$ indices in the expressions on the right hand side. No ${\cal Q}_X \leftrightarrow {\cal Q}_X^\dagger$ interchange is needed, unlike in the case of the amplitudes.

\subsection{Production}\label{subsec:production-amps}

\subsubsection{Pion decay}

Neutrino production through pion decay is elucidated in Fig.~\ref{fig:feynman-diagrams}. The corresponding decay amplitudes are 
\begin{eqnarray}
   \left.\mathcal{M}^\pi_{\mu k}\right\vert_{\nu_L} &\equiv& {\cal M}(\pi^+ \to \nu_{k,L} \,\mu^+) = - i\,m_\mu f_\pi\frac{V_{ud}}{v^2}\left[U^\dagger \cal P^\dagger \right]_{k\mu}(\bar{u}_{\nu_k} P_R v_{\mu})~, \nn
   \left.\mathcal{M}^\pi_{\mu k}\right\vert_{\nu_R} &\equiv& {\cal M}(\pi^+ \to \nu_{k,R} \,\mu^+)=-i\,m_\mu f_\pi\frac{V_{ud}}{v^2}\left[U^\dagger \widetilde{\cal P}^\dagger\right]_{k \mu}(\bar{u}_{\nu_k} P_L v_{\mu})~,
\end{eqnarray}
where the pion decay constant $f_{\pi}$ is defined by $\langle 0 | \bar d(0) \gamma_\mu \gamma_5 u(0) |\pi^+(p) \rangle   =   i \,p_\mu f_{\pi}$ and $\cal{P}$ and $\widetilde{\cal{P}}$ are combinations of the Wilson coefficients in Eq.~\eqref{eq.Lag-pion-decay}, namely
\begin{align}
    \mathcal{P}_{\alpha\beta} &= \left[\mathbb{I}+\epsilon^{\,ud}_L-\epsilon^{\,ud}_R\right]_{\alpha\beta} - \left[\epsilon^{\,ud}_P\right]_{\alpha\beta}\,\cfrac{m^2_{\pi^\pm}}{m_{l_\alpha}(m_u+m_d)}~,  \nn
    \widetilde{\cal{P}}_{\alpha\beta} &= \left[\widetilde{\epsilon}^{\,ud}_L-\widetilde{\epsilon}^{\,ud}_R\right]_{\alpha\beta} -\left[\widetilde{\epsilon}^{\,ud}_P\right]_{\alpha\beta}\,\cfrac{m^2_{\pi^\pm}}{m_{l_\alpha}(m_u+m_d)}~.
\end{align}
The integrated squared amplitudes are
\begin{eqnarray}\label{eq.AmpPiLH}
    \left.\int d\Pi_{\pi'}\mathcal{M}^\pi_{\mu k}\Bar{\mathcal{M}}^\pi_{\mu l}\right\vert_{\nu_L} &=& 2m_\pi\Gamma_\pi\frac{\left[U^\dagger {\cal P}^\dagger\right]_{k \mu} \left[\mathcal{P} U \right]_{\mu l} }{\left[\cal P\cal P^\dagger + \widetilde{\cal P}\widetilde{\cal P}^\dagger\right]_{\mu \mu}}\delta(E_\nu-E_{\nu,\pi})
    \nn
\label{eq.AmpPiRH}
    \left.\int d\Pi_{\pi'}\mathcal{M}^\pi_{\mu k}\Bar{\mathcal{M}}^\pi_{\mu l}\right\vert_{\nu_R} &=& 2m_\pi\Gamma_\pi\frac{\left[U^\dagger \widetilde{\cal P}^\dagger \right]_{k\mu} \left[\widetilde{\cal P} U\right]_{\mu l} }{\left[\cal P\cal P^\dagger + \widetilde{\cal P}\widetilde{\cal P}^\dagger\right]_{\mu \mu}}\,\delta(E_\nu-E_{\nu,\pi}),
\end{eqnarray}
where $E_{\nu,\pi}=(m_{\pi^\pm}^2-m_\mu^2)/(2m_{\pi^\pm})$ is the fixed energy of the neutrino in the decay. 
The $f_\pi^2\,V_{ud}^2/v^4$ factor is traded by the experimental pion decay width, $\Gamma_\pi$,  which introduces New Physics (NP) effects parametrized by the $1/\left[\cal P\cal P^\dagger + \widetilde{\cal P}\widetilde{\cal P}^\dagger\right]_{\mu \mu}$ factor.

\subsubsection{Muon decay}
The amplitudes for the neutrino and antineutrino production from muon decay, see Fig.~\ref{fig:feynman-diagrams}, are
\begin{align}\label{eq:muon-decay-amplitude}
    \left.\mathcal{M}^\mu_{j k}\right\vert_{\nu_H} &\equiv {\cal M}(\mu^+ \to \bar \nu_{j} e^+ \nu_{k,H} ) = -\frac{2}{v^2}\,\sum_{X,\eta}\,[U^\dagger(h^{X}_{\epsilon\eta})^\dagger U]_{kj}(\bar{v}_{\mu_{\eta}}\Gamma^X v_{\nu_{\rho}})(\bar{u}_{\nu_{H}} \Gamma_X v_{e_\epsilon}),  \nn %&& \\
    \left.\mathcal{M}^\mu_{j k}\right\vert_{\bar\nu_H} &\equiv {\cal M}(\mu^+ \to \bar \nu_{j,H} e^+ \nu_{k} ) = -\frac{2}{v^2}\,\sum_{X,\epsilon}\,[U^\dagger(h^{X}_{\epsilon\eta})^\dagger U]_{kj}(\bar{v}_{\mu_{\eta}}\Gamma^X v_{\nu_{H}})(\bar{u}_{\nu_{\rho}} \Gamma_X v_{e_\epsilon}),  %\nn &&
\end{align}
where $X=S,V,T$ refers to the Lorentz structure, $H=L,R$ is the chirality of (anti)neutrino, and $\eta=L,R$ ($\epsilon=L,R$) determines the chirality of the muon (electron) in the first (second) equation. The remaining chiralities are fixed by $X,H$ and $\eta(\epsilon)$.
The integrated squared amplitudes can be written compactly as
\begin{eqnarray}\label{eq.AmpMuLH}
    \sum_j\int d\Pi_{\mu'}\mathcal{M}^\mu_{j k}\Bar{\mathcal{M}}^\mu_{j l}\bigg\vert_x &=& 384\,\Gamma_\mu
    \frac{E_\nu^2}{m_\mu^2}
    \bigg[U^\dagger \left\{\left(\frac{1}{2}-\frac{E_\nu}{m_\mu}\right) \mathcal{H}_x^{(1)} + \frac{1}{3}\left(\frac{3}{4}-\frac{E_\nu}{m_\mu}\right) \mathcal{ H}_x^{(2)}
    \right\} U\bigg]_{kl}~~~~~~
\end{eqnarray}
where $x=\nu_L, \nu_R, \bar{\nu}_L, \bar{\nu}_R$ describes the neutrino species, $\Gamma_\mu = m_\mu^5\, / ( 384\pi^3\,v^4 )$ denotes the muon decay width, and $\cO(m_e/m_\mu)$ contributions have been neglected. For left-handed (anti)neutrinos, the $\mathcal{H}_x^{(i)}$ factors are given by
\begin{eqnarray}\label{eq:T-i}
    {\cal H}^{(1)}_{\nu_L} &=& h^{V\,\dagger}_{LL} \, h^{V}_{LL} 
    + 2 h^{T\,\dagger}_{RL} \, h^{T}_{RL} - \frac{1}{2}\left(h^{S\,\dagger}_{RL} \, h^{T}_{RL} + h^{T\,\dagger}_{RL} \, h^{S}_{RL}\right)\,, 
    \nn
    {\cal H}^{(2)}_{\nu_L} &=& h^{V\,\dagger}_{LR} \, h^{V}_{LR} + \frac{1}{4} h^{S\,\dagger}_{RR} \, h^{S}_{RR} + \frac{1}{4} h^{S\,\dagger}_{RL} \, h^{S}_{RL} + h^{T\,\dagger}_{RL} \, h^{T}_{RL} + \frac{1}{2}\left( h^{S\,\dagger}_{RL} \, h^{T}_{RL} +  h^{T\,\dagger}_{RL} \,h^{S}_{RL} \right)\,, \nn
    {\cal H}^{(1)}_{\bar\nu_L} &=& \frac{1}{4} h^{S}_{RR}  \, h^{S\,\dagger}_{RR} + \frac{1}{4} h^{S}_{LR} \, h^{S\,\dagger}_{LR} - h^{T}_{LR} \, h^{T\,\dagger}_{LR} \,,
    \nn
    {\cal H}^{(2)}_{\bar\nu_L} 
    &=&
    h^{V}_{LL} \, h^{V\,\dagger}_{LL} + h^{V}_{RL} \, h^{V\,\dagger}_{RL} + 4\, h^{T}_{LR} \,h^{T\,\dagger}_{LR}\,.
\end{eqnarray}

\begin{figure}
    \centering
    \includegraphics[width=0.45\linewidth]{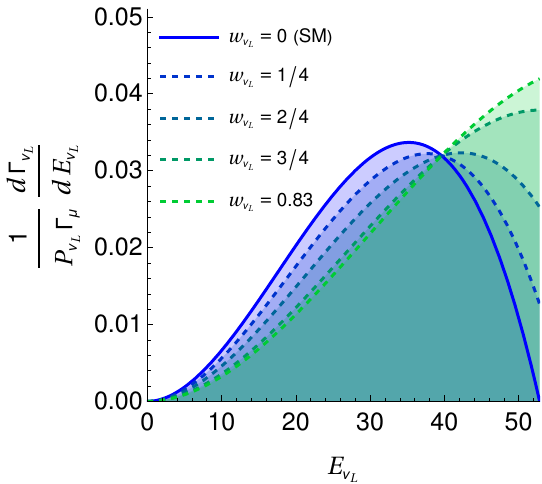}
    \hspace{0.5cm}
    \includegraphics[width=0.45\linewidth]{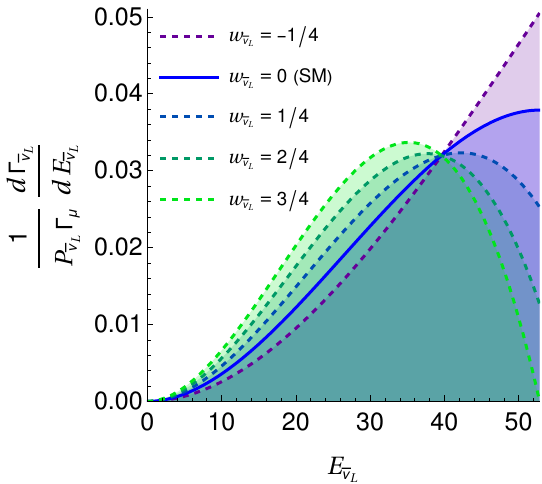}
    \caption{Neutrino (left) and antineutrino (right) spectrum for different values of the $w_{\nu_{{}_{L}}}$ and  $w_{\bar\nu_{{}_{L}}}$ parameters respectively. The case $w_{\nu_{{}_{L}}} = w_{\bar\nu_{{}_{L}}} = 0$ corresponds to the  SM scenario.}
    \label{fig:MuonDecaySpectrum}
\end{figure}

From Eq.~\eqref{eq.AmpMuLH}, we can obtain the differential muon-decay rate 
with respect to neutrino and anti-neutrino energies. This is achieved by setting $k = l$ and summing over $k$. This leads us to the following parametric form
\begin{eqnarray}
\cfrac{d\,\Gamma_{\nu_L}}{d E_{\nu_{_L}}}&= \cfrac{24\,\Gamma_\mu}{m_\mu}\, y^2\, P_{\nu_{_L}}\bigg[ \left(1-y\right) +\,\cfrac{8}{9}\,  w_{\nu_{_L}} \left(y-\cfrac{3}{4}\right)
\bigg]~,\nn
\label{eq:AmpMuLH-PDGnotation-nubar}
\cfrac{d\,\Gamma_{\bar\nu_L}}{d E_{\bar\nu_{_L}}}&= \cfrac{24\,\Gamma_\mu}{m_\mu}\, y^2\,P_{\bar\nu_{_L}} \bigg[\left(\cfrac{1}{2}-\cfrac{y}{3}\right) +\,\cfrac{8}{9}\,  w_{\bar{\nu}_{_L}} \left(\cfrac{3}{4} - y \right)\bigg]~,
\end{eqnarray}
for neutrinos and antineutrinos respectively, where
$y=2E_\nu / m_\mu$\footnote{We use $ E_{\nu_{_L}} = E_{\bar\nu_{_L}} = E_\nu $.} and
\begin{align}
    P_{\nu_L} &= {\rm Tr} \left[ \mathcal{ H}_{\nu_L}^{(1)}\,+\,\mathcal{ H}_{\nu_L}^{(2)}\right] = {\rm Tr} \left[ |h^V_{LL}|^2 + |h^V_{LR}|^2 + \frac{1}{4}|h^S_{RR}|^2 + \frac{1}{4}|h^S_{RL}|^2 + 3|h^T_{RL}|^2 \right],\nn
    P_{\bar\nu_L} &= {\rm Tr} \left[ \mathcal{ H}_{\bar\nu_L}^{(1)}\,+\,\mathcal{ H}_{\bar\nu_L}^{(2)}\right] = {\rm Tr} \left[ |h^V_{LL}|^2 + |h^V_{RL}|^2 + \frac{1}{4}|h^S_{RR}|^2 + \frac{1}{4}|h^S_{LR}|^2 + 3|h^T_{LR}|^2 \right]~,\nn
    w_{\nu_{{}_{L}}} &= \frac{3}{4}\, \frac{{\rm Tr} \left[\mathcal{H}_{\nu_{{}_{L}}}^{(2)}\right]}{P_{\nu_{{}_{L}}}} = 
    \frac{3}{4}\frac{{\rm Tr} \left[ 4 |h^{V}_{LR}|^2 +  |h^{S}_{RR}|^2 + \left| h^{S}_{RL} + 2 h^{T}_{RL} \right|^2 \right]}{{\rm Tr} \left[ 4 |h^V_{LL}|^2 + 4\, |h^V_{LR}|^2 + |h^S_{RR}|^2 + |h^S_{RL}|^2 + 12|h^T_{RL}|^2  \right]}~, \nn 
    w_{\bar\nu_{{}_{L}}} &= \frac{3}{4}\, \frac{{\rm Tr} \left[\mathcal{H}_{\bar\nu_{{}_{L}}}^{(1)}\right]}{P_{\bar\nu_{{}_{L}}}} = 
    \frac{3}{4}\frac{{\rm Tr} \left[ |h^{S}_{RR}|^2 + |h^{S}_{LR}|^2 - 4 |h^{T}_{LR}|^2 \right]}{{\rm Tr} \left[ 4 |h^V_{LL}|^2 + 4 |h^V_{RL}|^2 + |h^S_{RR}|^2 + |h^S_{LR}|^2 + 12 |h^T_{LR}|^2 \right] }~.
    \label{eq:Q-prob-muon-decay}
\end{align} 
Here, we have introduced the notation $|A|^2 = A^\dagger A$ for brevity. $P_{\nu_L}$ and $P_{\bar\nu_L}$ denote the overall probability for the production of a left-handed neutrino and an anti-neutrino respectively. Finally,  $w_{\nu_{{}_{L}}}$ and  $w_{\bar\nu_{{}_{L}}}$ describe the spectral shape parameters for neutrinos and anti-neutrinos respectively\footnote{The allowed values for $P_{\nu_H}w_{\nu_H}$ and $P_{\bar\nu_H}w_{\bar\nu_H}$ are $[0,0.83]$ and $[-\frac{1}{4},\frac{3}{4}]$ respectively. In the first case the maximum value is obtained for $h^V_{LR} = h^S_{RR} = 0$, $|h^S_{RL}| = 1.87$, $|h^T_{RL}| = 0.20$ and no interference angle. In the second case the maximum is obtained for $|h^{S}_{RR}|^2+|h^{S}_{LR}|^2 = 4$ and the minimum for $|h^{T}_{LR}|^2 = 1/3$.},  as shown in Fig.~\ref{fig:MuonDecaySpectrum}. 

The $\mathcal{H}_x^{(i)}$ matrices, differential decay widths and the associated $P_x$ and $w_x$ parameters for the production of right-handed (anti)neutrinos can be obtained from Eqs.~\eqref{eq:T-i}-\eqref{eq:Q-prob-muon-decay} 
by flipping the chirality indices $L\leftrightarrow R$.

Eqs.~\eqref{eq:AmpMuLH-PDGnotation-nubar}-\eqref{eq:Q-prob-muon-decay}  
generalize the results presented in Ref.~\cite{Fetscher:1994nv}\footnote{We note that there is a $-8/9$ factor missing in Eq.~(57.8) of the PDG review on muon-decay parameters~\cite{MuonDecayParameters:2024cfk}, i.e., in the neutrino energy distribution that we show in the first line of the Eq.~\eqref{eq:AmpMuLH-PDGnotation-nubar}.} for neutrinos and in our recent letter~\cite{Breso-Pla:2025cul} for antineutrinos when lepton-flavor conservation is not imposed. In other words, the flavor of the emitted neutrino (antineutrino) is generic, i.e., not necessarily $\nu_e$ ($\bar\nu_\mu$). It is worth stressing that the parametric form of the differential decay widths, Eq.~\eqref{eq:AmpMuLH-PDGnotation-nubar}, is unaltered by this generalization, which only affects the definition of the $P_x$ and $w_x$ parameters. 
In the SM, all $P_x$ and $w_x$ are zero except for $P_{\nu_L}=P_{\bar\nu_L}=1$.

The normalization condition for the muon-decay Wilson Coefficients, shown in Eq.~\eqref{eq:muon_decay_norm}, can be rewritten as
\begin{eqnarray}
\label{eq:normalizationBIS}
    {\rm Tr}\left[\mathcal{ H}_{\nu_L}^{(1)} + \mathcal{ H}_{\nu_L}^{(2)} + \mathcal{ H}_{\nu_R}^{(1)} + \mathcal{ H}_{\nu_R}^{(2)}\right] \,&=&\, P_{\nu_L} + P_{\nu_R}  =  1~, \nn
    {\rm Tr}\left[\mathcal{ H}_{\overline{\nu}_L}^{(1)} + \mathcal{ H}_{\overline{\nu}_L}^{(2)} + \mathcal{ H}_{\overline{\nu}_R}^{(1)} + \mathcal{ H}_{\overline{\nu}_R}^{(2)}\right] \,&=&\, P_{\bar\nu_L} + P_{\bar\nu_R}  = 1~.
\end{eqnarray}

\section{Event rate}\label{sec:rates}
Using the integrated squared amplitudes derived in the previous section we can calculate the event rates mediated by neutrinos produced in pion decays and by (anti)neutrinos produced in muon decays, {\it cf.} Eq.~(\ref{eq:DifferentialRateEvents}). 
Summing the three contributions and integrating over the neutrino energy we obtain the following rate per recoil energy $T$ and time $t$
    \begin{eqnarray}
    \label{eq:rate}
    \frac{dN}{dtdT}=g_\pi(t)\frac{dN^{\rm prompt}}{dT}+g_\mu(t)\frac{dN^{\rm delayed}}{dT}~,
    \end{eqnarray}
with
    \begin{eqnarray}
        \label{eq:prompt_and_delayed}
        \hspace*{-0.5cm} \frac{dN^{\rm prompt}}{dT} &=& \cfrac{N_\pi N_T \mathcal{F}(q^2)^2m_\mathcal{N}}{32\pi^2v^4L^2} \left\{(\widetilde{Q}_{_V}^\mu)^2 f_{_V}^{\mu}(T)+
        \hspace{-0.1cm}\sum_{X=S,F,SF}\hspace{-0.1cm}(\widetilde{Q}_{_X}^\mu)^2 f_{_X}^\mu(T)\right\},\nn
        \hspace*{-0.5cm} \frac{dN^{\rm delayed}}{dT} &=& \cfrac{N_\mu N_T \mathcal{F}(q^2)^2m_\mathcal{N}}{32\pi^2v^4L^2} \left\{(\widetilde{Q}_{_V}^{\bar{\mu}})^2 f_{_V}^{\bar{\mu}}(T)+(\widetilde{Q}_{_V}^e)^2 f_{_V}^{e}(T) + \hspace{-0.1cm}\sum_{\substack{X=S,F,SF \\ \alpha=\bar\mu,e}} \hspace{-0.2cm}(\widetilde{Q}_{_X}^\alpha)^2 f_{_X}^\alpha(T)\right\},
    \end{eqnarray}
where $f_i^\alpha(T)$ are kinematic functions (defined in App. \ref{app:fullresults}) and the generalized charges $\widetilde{Q}_i^\alpha$ are functions of both production and detection Wilson Coefficients, which we discuss later in this section. We have singled out the vector terms because these are the only terms generated in the SM.
Much as in Ref.~\cite{Breso-Pla:2023tnz}, the rate can be conveniently written as SM fluxes times effective differential cross sections:
\begin{eqnarray}
    \label{eq:prompt_and_delayedBIS}
    \frac{dN^{\rm prompt}}{dT} &=& N_T\int dE_\nu\frac{d\phi_{\nu_\mu}^{\rm SM}}{dE_\nu}\frac{d\,\widetilde{\sigma}_{\nu_\mu}}{d\,T}\nn
    \frac{dN^{\rm delayed}}{dT} &=& N_T\int dE_\nu\left(\frac{d\phi_{\nu_e}^{\rm SM}}{dE_\nu}\frac{d\,\widetilde{\sigma}_{\nu_e}}{d\,T}+\frac{d\phi_{\bar\nu_\mu}^{\rm SM}}{dE_\nu}\frac{d\,\widetilde{\sigma}_{\bar\nu_\mu}}{d\,T}\right)
\end{eqnarray}
where the (effective) differential cross sections are given by
\begin{eqnarray}\label{eq.crosssec}
    \frac{d\,\widetilde{\sigma}_{f}}{dT} &=&\frac{m_\mathcal{N}\mathcal{F}^2(q^2)}{8\pi v^4}\left\{(\widetilde{Q}_{_V}^f)^2\left(1-\frac{m_\mathcal{N}T}{2E^2_\nu}-\frac{T}{E_\nu}\right) +(\widetilde{Q}_{_S}^f)^2\,\frac{m_\mathcal{N} T}{2E^2_\nu} \right.\nn
    && \qquad\qquad ~~+\left.(\widetilde{Q}_{_F}^f)^2\frac{v^2}{2m_\mathcal{N} T}\left(1-\frac{T}{E_\nu}\right) +(\widetilde{Q}_{_{SF}}^f)^2\frac{v}{2E_\nu}\left(1-\frac{T}{2E_\nu}\right)\right\},
\end{eqnarray}
where $f=\mu,~\bar\mu,~e$ and $0\leq T \leq 2E_{\nu}^2/(m_{\cal N}+2E_{\nu})$. The SM fluxes are given by their usual expressions:
\begin{align}\label{eq:SMfluxes}
\frac{d\phi^{\rm SM}_{\nu_\mu}}{dE_\nu} &= \frac{N_{\nu_\mu}}{4 \pi L^{2}} \delta (E_\nu - E_{\nu,\pi})~,\nn   
\frac{d\phi^{\rm SM}_{\nu_e}}{dE_\nu} &= \frac{N_{\nu_e}}{4 \pi L^{2}} \frac{192 E_{\nu}^2}{m_{\mu}^3}\left(\frac{1}{2}-\frac{E_{\nu}}{m_{\mu}} \right)~,\nn    
\frac{d\phi^{\rm SM}_{\bar{\nu}_\mu}}{dE_\nu} &= \frac{N_{\bar{\nu}_\mu}}{4 \pi L^{2}} \frac{64 E_{\nu}^2}{m_{\mu}^3}\left(\frac{3}{4}-\frac{E_{\nu}}{m_{\mu}} \right)~,
\end{align}
where $0\leq E_\nu\leq m_\mu/2$ for muon-decay fluxes, $N_{\nu_\mu} = n_{\rm POT}\, r_{\pi/p}\, BR(\pi\to \mu\nu)_{\rm exp}$, and $N_{\bar{\nu}_\mu} = N_{\nu_e} = n_{\rm POT}\, r_{\mu/p}\, BR(\mu\to e\bar{\nu}\nu)_{\rm exp}$ are the total number of (anti)neutrinos emitted through pion and muon decays respectively.

The generalized charges $\widetilde{Q}_{_{V,S,F,SF}}^f$ are lengthy functions of the production and detection Wilson Coefficients, which can be written in the following compact form
\begin{eqnarray}\label{eq:Qsq-full-exprs}
    (\widetilde{Q}_{X}^\mu)^2  &=& \cfrac{\left[{\cal P}A_{X}^2 {\cal P}^\dagger+\widetilde{\cal P}A_{\widetilde X}^2 \widetilde{\cal P}^\dagger \right]_{\mu\mu}}{\left[{\cal P}{\cal P}^\dagger +\widetilde{\cal P}\widetilde{\cal P}^\dagger\right]_{\mu\mu}} ~,\nn
(\widetilde{Q}_{X}^{\bar{\mu}})^2 
&=& {\rm Tr}\left[\left({\cal H}^{(2)}_{\nu_L} + {\cal H}^{(2)}_{\bar{\nu}_L}\right)\,A_X^2 + \left({\cal H}^{(2)}_{\nu_R} + {\cal H}^{(2)}_{\bar{\nu}_R}\right) A_{\widetilde X}^2\right]~, \nn
(\widetilde{Q}_{X}^{e})^2 
&=& {\rm Tr}\left[\left({\cal H}^{(1)}_{\nu_L} + {\cal H}^{(1)}_{\bar{\nu}_L}\right)\,A_X^2 + \left({\cal H}^{(1)}_{\nu_R} + {\cal H}^{(1)}_{\bar{\nu}_R}\right)\,A_{\widetilde X}^2\right]~,
\end{eqnarray}
with 
$A_X^2 = \{{\cal Q}_V^2,\,{\cal Q}_{S}{\cal Q}_{S}^\dagger,\, {\cal Q}_F{\cal Q}_F^\dagger, \, {\cal Q}_F{\cal Q}_S^\dagger+{\cal Q}_S{\cal Q}_F^\dagger\}$ 
and 
$A_{\widetilde X}^2 = \{{\cal Q}_{\widetilde V}^2,\,{\cal Q}_S^\dagger {\cal Q}_S,\, {\cal Q}_F^\dagger {\cal Q}_F,\, {\cal Q}_S^\dagger {\cal Q}_F + {\cal Q}_F^\dagger {\cal Q}_S\}$
for $X\,=\,\{V,\,S,\,F,\,SF\}$. The $(\widetilde{Q}_{_X}^f)^2$ quantities are real but not necessarily positive (see, e.g., the $SF$ interference term), but our notation allows us to write the results in a compact form.

The generalized vector charges,  $(\widetilde{Q}_{V}^{\mu})^2$, $(\widetilde{Q}_{V}^{\bar{\mu}})^2$ and $(\widetilde{Q}_{V}^{e})^2$, are the only ones that are not zero in the SM limit. Because of that, it is interesting to single out the SM contribution, and to rewrite these charges as follows
\begin{align}\label{eq:Qvec-deviation-from-SM}
    (\widetilde{Q}_V^f)^2 &= Q_{\text{SM}}^2\,x_f 
    + \Delta_f~,
\end{align}
where the $x_f$
factors, which include only non-standard production effects, are given by 
\begin{align}\label{eq:xdefinitions}
x_\mu &=  \cfrac{ \left[{\cal P}{\cal P}^\dagger \right]_{\mu\mu}}{\left[{\cal P}{\cal P}^\dagger +\widetilde{\cal P}\widetilde{\cal P}^\dagger\right]_{\mu\mu}} ~, \nn
x_{\bar{\mu}} &=  \,{\rm Tr}\left[{\cal H}^{(2)}_{\nu_L} + {\cal H}^{(2)}_{\bar\nu_L} \right] =  P_{\bar\nu_L} - \frac{4}{3}\,P_{\bar\nu_L} w_{\bar\nu_L} + \frac{4}{3}\,P_{\nu_L} w_{\nu_L}~, \nn
x_e &= \,{\rm Tr}\left[{\cal H}^{(1)}_{\nu_L} + {\cal H}^{(1)}_{\bar\nu_L} \right] = P_{\nu_L} - \frac{4}{3}\,P_{\nu_L} w_{\nu_L}  + \frac{4}{3}\,P_{\bar\nu_L} w_{\bar\nu_L}~,
\end{align}
and the $\Delta_f$ factors, which encode non-standard detection effects, are given by
\begin{eqnarray}
\Delta_\mu &=& \cfrac{\left[{\cal P}\Delta_V {\cal P}^\dagger + \widetilde{\cal P}Q_{\widetilde V}^2 \widetilde{\cal P}^\dagger \right]_{\mu\mu}}{\left[{\cal P}{\cal P}^\dagger +\widetilde{\cal P}\widetilde{\cal P}^\dagger\right]_{\mu\mu}}~, 
\nn
\Delta_{\bar\mu} 
&=& {\rm Tr}\left[\left({\cal H}^{(2)}_{\nu_L} + {\cal H}^{(2)}_{\bar{\nu}_L}\right)\,\Delta_V + \left({\cal H}^{(2)}_{\nu_R} + {\cal H}^{(2)}_{\bar{\nu}_R}\right)\,{\cal Q}_{\widetilde V}^2\right]~, \nn
\nn
\Delta_e
&=& {\rm Tr}\left[\left({\cal H}^{(1)}_{\nu_L} + {\cal H}^{(1)}_{\bar{\nu}_L}\right)\,\Delta_V + \left({\cal H}^{(1)}_{\nu_R} + {\cal H}^{(1)}_{\bar{\nu}_R}\right)\,{\cal Q}_{\widetilde V}^2\right]~,
\end{eqnarray}
where $\Delta_V \equiv\,{\cal Q}_V^2 - Q_{\text{SM}}^2\mathbb{I}$ is the deviation from the SM value.

Let us make some final remarks. 
First, we note that the PMNS matrix does not appear in the final result, as it should be since $L\approx 0$. 
Secondly, if we drop all operators with right-handed neutrinos, only the vector charges remain and we recover for them the same expressions as in Ref.~\cite{Breso-Pla:2023tnz}. As pointed out in that work, that limit reproduces the SM rate but with weak nuclear charges for $\nu_\mu, \nu_e$, and $\bar\nu_\mu$ containing non-standard effects affecting detection and production.
It is well known that this changes with the introduction of operators with right-handed neutrinos, since the scalar interactions and the neutrino magnetic moments affect the cross section in a novel form, namely, with a non-standard $T$ dependence~\cite{Lindner:2016wff,Vogel:1989iv}.

Our results above include for the first time non-standard interactions with right-handed neutrinos also in the production processes (pion and muon decay). 
The parametrization in Eq.~(\ref{eq:prompt_and_delayedBIS}), which is convenient from a practical point of view, hides this fact, since it uses the SM fluxes and moves the non-standard production effects to the {\it effective} cross sections $\tilde\sigma_f$. 

Even in the absence of lepton-flavor violation, that is, when the (anti)neutrinos have the same flavor as in the SM, the delayed neutrino flux is not proportional to its SM value, and likewise for the delayed antineutrino flux~\cite{Breso-Pla:2025cul}. This is clear from the comparison of Eq.~\eqref{eq:AmpMuLH-PDGnotation-nubar} and Eq.~\eqref{eq:SMfluxes}, due to the $w_x$ terms. Only the sum of the neutrino and antineutrino contributions to the event rate can be written in the form of SM fluxes times effective differential cross sections, as in Eq.~\eqref{eq:prompt_and_delayedBIS}. 

If lepton-flavor violation is allowed one can have, for instance, tau-neutrinos contributing to the event rate. This is fully taken into account in the practical parametrization in Eq.~\eqref{eq:prompt_and_delayedBIS} despite not having a $\nu_\tau$ flux. Thus, one should keep in mind that the subindices $\nu_e,\nu_\mu$ and $\bar\nu_\mu$ do not have physical meaning in general, as discussed in Ref.~\cite{Breso-Pla:2023tnz}.

Let us now discuss the form of our results in some interesting limits.

\subsection{NP up to second order}\label{subsec:NP-upto-2nd-order}

The general expressions for the generalized charges become much simpler if we neglect terms beyond second order in dim-6 $\nu$WEFT Wilson Coefficients, namely
\begin{align}
    (\widetilde{Q}_{_S}^f)^2 &= \left[ {\cal Q}_S {\cal Q}_S^\dagger \right]_{ff}~,\nn
    (\widetilde{Q}_{_F}^f)^2 &= \left[{\cal Q}_F {\cal Q}_F^\dagger\right]_{ff}~,\nn
    (\widetilde{Q}_{_{SF}}^f)^2 &= \left[{\cal Q}_S {\cal Q}_F^\dagger + {\cal Q}_F {\cal Q}_S^\dagger\right]_{ff}~,
\end{align}
for $f=e,\mu$, whereas $\widetilde{Q}_X^{\bar{\mu}}=\widetilde{Q}_X^\mu$ for $X=S,F,SF$, i.e., equal for muon neutrinos and  antineutrinos. 
For the vector charges we have
\begin{align}
    \label{eq:vectorexpansion}
    (\widetilde{Q}_{_V}^\mu)^2 &= Q_{\text{SM}}^2 \left(1 - \left[\widetilde{\cal P}\widetilde{\cal P}^\dagger\right]_{\mu\mu} \right) \,+\, \cfrac{ \left[{\cal P} \,\Delta_V\, {\cal P}^\dagger \right]_{\mu\mu}}{\left[{\cal P}{\cal P}^\dagger\right]_{\mu\mu}}
    ~,\nn
    (\widetilde{Q}_{_V}^{\bar{\mu}})^2 &= Q_{SM}^2\, x_{\bar{\mu}}  \,+\, {\rm Tr}[h^{V\,\dagger}_{LL} \,\Delta_V\, h^{V}_{LL}] ~,\nn
    (\widetilde{Q}_{_V}^{e})^2 &= Q_{SM}^2\, x_e \,+\, {\rm Tr}[h^{V}_{LL} \,\Delta_V\, h^{V\,\dagger}_{LL}]~.
\end{align}
We do not expand the second term in Eqs.~(\ref{eq:vectorexpansion}) because it would lead to less compact expressions, but we note that these terms contain the only correction that is linear in NP, which was studied in detail in Ref.~\cite{Breso-Pla:2023tnz}.

Overall, we go from 12 parameters in the most general case to 9 parameters after this expansion, namely $\widetilde{Q}_{_{S,F,SF}}^f$ ($f=\mu,e$), $\widetilde{Q}_{_{V}}^\mu$, $\widetilde{Q}_{_{V}}^{\bar\mu}$ and $\widetilde{Q}_{_{V}}^e$.

\subsection{NP only in detection and comparison with previous works}\label{subsec:NP-only-in-detection}

In order to compare our results with previous works, we assume NP only in detection, i.e.,
\begin{eqnarray}
\left[\cal{P}\right]_{\alpha \beta}&\longrightarrow&\delta_{\alpha \beta}, \qquad
\left[h_{LL}^{V}\right]_{\alpha \beta} \longrightarrow \delta_{\alpha \mu}\delta_{\beta e}~,
\end{eqnarray}
and set the remaining coefficients to zero. This does not change the parametric result for the observable, \textit{cf.} Eqs.~(\ref{eq:rate})-(\ref{eq:prompt_and_delayedBIS}), but it affects the underlying meaning of the generalized charges, which now take the following form
\begin{eqnarray}\label{eq.CNPoDPi}
(\widetilde{Q}_{_V}^f)^2 &=& \left[{\cal Q}_V^2\right]_{ff}~,\qquad\qquad (\widetilde{Q}_{_S}^f)^2 = \left[{\cal Q}_S {\cal Q}_S^\dagger\right]_{ff}~,\nn
(\widetilde{Q}_{_F}^f)^2 &=& \left[{\cal Q}_F {\cal Q}_F^\dagger\right]_{ff}~, \qquad
(\widetilde{Q}_{_{SF}}^f)^2 = \left[{\cal Q}_S {\cal Q}_F^\dagger + {\cal Q}_F {\cal Q}_S^\dagger\right]_{ff}~,
\end{eqnarray}
for $f=e,\mu$. In this case, all generalized charges of the muon neutrinos and antineutrinos are equal, including the vector ones, i.e., $\widetilde{Q}_{_X}^{\bar\mu}=\widetilde{Q}_{_X}^{\mu}$, $X=V,S,F,SF$.  
Thus, one has 8 parameters, namely $\widetilde{Q}_{_{V,S,F,SF}}^f$ for $f=\mu,e$. Let us also note that in this case, since the production occurs through SM interactions, the factorization in flux times cross section is not just a convenient parametrization but an actual physical result. For instance, the generalized charges in the cross section depend only on detection physics and there are no $\nu_\tau$-mediated contributions.

Let us now compare our results with previous literature:
\begin{itemize}
    \item In the SM limit we recover the familiar expression for the CE$\nu$NS cross section with only the vector charge~\cite{Freedman:1973yd}, which is flavor universal up to negligible radiative corrections~\cite{Tomalak:2020zfh}. 
    \item In the presence of non-standard interactions involving only left-handed neutrinos, we also recover the well-known expression, with one vector charge for the $\nu_e$ channel and another one for the $\nu_\mu$ and $\bar\nu_\mu$ channels~\cite{Barranco:2005yy,Scholberg:2005qs,Lindner:2016wff,Breso-Pla:2023tnz}. 
    \item If vector interactions with right-handed neutrinos are added, our results agree with Refs.~\cite{Lindner:2016wff} with the correspondence $(\widetilde{Q}_{_V}^f)^2\to  N^2\xi_V^2$, where $N=A-Z$ is the number of neutrons in the nucleus, and assuming lepton universality\footnote{For a spin-$1/2$ nucleus, Ref.~\cite{Lindner:2016wff} finds $N^2\xi_V^2 =(C_V- D_A)^2$, which we confirm, contrary to the expression in Ref.~\cite{AristizabalSierra:2018eqm}. $C_V$ and $D_A$ are the nucleus-neutrino couplings, which are connected with quark-neutrino couplings by $C_V = Z(2C^u_V+C^d_V)+(A-Z)(C^u_V+2C^d_V)$ and $D_A = Z(2 D^u_A+ D^d_A)+(A-Z)( D^u_A+2 D^d_A)$. In our notation: 
    $C_V^q \equiv g_V^{qq}+[\epsilon_V^{qq}+\widetilde{\epsilon}_V^{\,qq}]_{\ell\ell}$, and $D_A^{q} \equiv -g_V^{qq}-[\epsilon_V^{qq}-\widetilde{\epsilon}_V^{\,qq}]_{\ell\ell}$.
    \label{footnote:xiV}
    }. Obviously vector interactions with right-handed neutrinos have no effect in the final observable in the absence of NP in detection.
   \item Scalar term: We agree with Refs.~\cite{Lindner:2016wff,AristizabalSierra:2018eqm} with the correspondence $(\widetilde{Q}_{_S}^f)^2\to  N^2\xi_S^2$ and the assumption of lepton universality and real couplings.
    \item Tensor term: As discussed in Sec.~\ref{subsec:nucleon-lepton-eft}, we agree with Refs.~\cite{Altmannshofer:2018xyo,Hoferichter:2020osn,Glick-Magid:2023uhk} but not with a long list of works, which included the tensor contribution among the coherently enhanced part of the cross section~\cite{Barranco:2011wx,Healey:2013vka,Papoulias:2015iga,Papoulias:2017qdn,AristizabalSierra:2018eqm,AristizabalSierra:2019zmy,Bischer:2019ttk,Papoulias:2019xaw,Han:2020pff,Li:2020lba,CONUS:2021dwh,Majumdar:2021vdw,AristizabalSierra:2022axl,Chen:2022xkk,Majumdar:2022nby,DeRomeri:2022twg,Fridell:2023rtr,Demirci:2023tui,Chatterjee:2024vkd,Karmakar:2024ywn}. As a result, the bounds on the tensor quark-level coefficients $\tilde{\epsilon}^{qq}_T$ obtained in these works are incorrect by a factor $N^2$.
    \item Magnetic moment: We agree with the previous literature, see e.g., Ref.~\cite{Vogel:1989iv,Giunti:2024gec}, which typically uses the so-called effective neutrino magnetic moment $\mu_{\nu_\ell}^2\equiv(\mu_\nu^\dagger \mu_\nu)_{\ell\ell}$. The relation with our notation is
    \begin{align}
    \left[ {\cal Q}_F^\dagger {\cal Q}_F\right]_{\ell\ell}=16\pi\alpha \,Z^2\,v^2\, \mu^2_{\nu_\ell} = Z^2\frac{16\pi^2\alpha^2\,v^2}{m_e^2} \frac{\mu^2_{\nu_\ell}}{\mu^2_B}~.
    \end{align}
    In addition to the primary CE$\nu$NS effect, the neutrino magnetic moment also contributes to the event rate through the (subleading) elastic neutrino-electron scattering. Phenomenological analyses indicate that this produces a small ($\sim 5\%$) improvement in the bounds~\cite{DeRomeri:2022twg}.
    \item Interference terms: Our calculation of the interference between scalar and magnetic-moment terms is in agreement with the expression presented in Ref.~\cite{Chang:2020jwl} for the one-flavor case (with the obvious replacement $\chi\to\nu$).
\end{itemize}

\subsection{NP only in production}\label{subsec:NP-only-in-production}
Let us discuss now the case where the non-standard effects only affect (anti)neutrino production. In that case the only non-zero nuclear charge is ${\cal Q}_V \rightarrow Q_{\text{SM}} \times \mathbb{I}$ and, as a consequence, the only non-zero generalized nuclear charges are the following
\begin{eqnarray}\label{eq:newQproduction}
(\widetilde{Q}_{_V}^\mu)^2  &=& Q_{\text{SM}}^2\, x_{\mu}~, \qquad
(\widetilde{Q}_{_V}^{\bar{\mu}})^2 = 
Q_{\text{SM}}^2\, x_{\bar{\mu}} ~, \qquad
(\widetilde{Q}_{_V}^{e})^2 = 
Q_{\text{SM}}^2\, x_e ~,
\end{eqnarray}
with the $x_f$ factors defined in Eq.~(\ref{eq:xdefinitions}).

First, let us discuss the results in the absence of right-handed neutrinos, i.e., $\widetilde{\cal P}=0$ and $h_{\epsilon\eta}^X=0$ except for $h_{LL}^V$ and $h_{RR}^S$. For the probabilities, we have $P_{\nu_L}=P_{\bar\nu_L}=1$, after taking into account the normalization condition, Eq.~(\ref{eq:muon_decay_norm}). The shape factors for left-handed (anti)neutrino differential distributions are not zero, namely, $P_{\nu_{{}_L}} w_{\nu_{{}_L}} = P_{\bar\nu_{{}_L}} w_{\bar\nu_{{}_L}} = 3\,{\rm Tr} |h_{RR}^S|^2/16$. 
However, they cancel in their contribution to the COHERENT observable, which cannot differentiate between neutrino and anti-neutrino events, i.e., $x_\mu=x_{\bar{\mu}}=x_e=1$. 
This corresponds to the scenario studied in Ref.~\cite{Breso-Pla:2023tnz}. Non-standard effects in production are entirely captured in the pion and muon decay widths (when NP in detection is absent), so COHERENT cannot be used to probe them.

The situation changes completely if interactions with right-handed neutrinos are included. Indeed, the generalized charges are functions of non-standard effects affecting pion decay and muon decay, {\it cf.}~Eq.~\eqref{eq:newQproduction}. Thus, COHERENT data can be used to constrain these effects. 
This point, which is one of our main results, was highlighted in the recent letter~\cite{Breso-Pla:2025cul} in the specific context of muon-decay parameters with lepton-flavor conservation.

In the case of pion decay, the
$x_\mu$ factor multiplying $Q_{\text{SM}}^2$ is simply the probability that the pion emits a left-handed neutrino of any flavor. In other words, it predicts the number of $\pi$-emitted neutrinos that contribute to COHERENT compared to the total number of neutrinos that are produced in pion decay. Equivalently, $1-x_\mu$ parametrizes the complementary $\nu_R$ fraction, which does not contribute to COHERENT. 

For muon decay, this distinction cannot be drawn so easily. As discussed in Ref.~\cite{Breso-Pla:2025cul}, the $\bar{\mu}$ and $e$ sub-indices in $x_{\bar\mu}$ and $x_e$ do not refer to the nature of the emitted particle, but to the parametric behavior in the event rate formula. Indeed the $x_{\bar\mu}$ term has contributions from antineutrinos (of any flavor) and also from neutrinos (of any flavor), and vice versa for the $x_e$ term. The sum of these factors does have a simple interpretation, namely, it corresponds to the probability that the muon emits a left-handed neutrino or a left-handed antineutrino: $x_{\bar{\mu}} + x_e = P_{\nu_L} + P_{\bar\nu_L}$.
The sensitivity of COHERENT data to the $x_f$ parameters was used in Ref.~\cite{Breso-Pla:2025cul} to extract for the first time a constraint on the $P_{\bar\nu_L}$ and $w_{\bar\nu_L}$ parameters that describe the antineutrino energy distribution.

Ultimately, the $x_f$ terms come from a comparison of the COHERENT rate with the experimental values of the pion and muon decay widths, $\Gamma_\pi$ and $\Gamma_\mu$, although this is less explicit in the case of the muon-decay due to the use of the normalized Wilson Coefficients $h^X_{\epsilon\eta}$ (instead of $\hat{h}^X_{\epsilon\eta}$), {\it cf.}~Eq.~\eqref{eq:Lag-muon-decay}.

\section{Phenomenological analysis}\label{sec:pheno}
\subsection{Experimental input}\label{subsec:exp-input}
We will use the two most precise available CE$\nu$NS measurements performed by the COHERENT collaboration: one using a liquid argon target (LAr)~\cite{COHERENT:2020iec} and another using a cesium iodide target (CsI)~\cite{COHERENT:2021xmm}. 

Taking into account various experimental effects~\cite{COHERENT:2020ybo, COHERENT:2021xmm}, the theoretical prediction for $dN^a/dT$ ($a={\rm prompt/delayed}$) given in Eq.~(\ref{eq:prompt_and_delayed}) is connected with the number of events detected in the $i$-th bin with the LAr detector by
\begin{eqnarray}
\label{eq:NumberofEventsWithExperimentalEffects}
N^a_{i} = \int_{T_{ee}^{\text{rec}, i}}^{T_{ee}^{\text{rec}, i+1}} dT_{ee}^{\text{rec}} \,\epsilon(T_{ee}^{\text{rec}}) \int_{T_{\text{min}}}^{T^a_{\text{max}}} dT \,{\cal R}(T_{ee}^{\text{rec}},T_{ee}(T)) \frac{dN^a}{dT}\left( T \right)\,.
\end{eqnarray}
The electron-equivalent recoil energy $T_{ee}$ is related to the nuclear recoil energy through the quenching factor, $T_{ee}(T)=\text{QF}(T)\times  T$. The energy resolution function, ${\cal R}(T_{ee}^{\text{rec}},T_{ee})$, relates the true value, $T_{ee}$, to the reconstructed one that is registered at the detector, $T_{ee}^{\text{rec}}$. Finally, $\epsilon(T_{ee}^{\text{rec}})$ is the detector efficiency. The expression in Eq.~\eqref{eq:NumberofEventsWithExperimentalEffects} also holds for CsI with the replacement $T_{ee}^{\text rec}\to$ PE (number of photoelectrons). The light yield LY (number of PE produced by an electron recoil of one keV) connects these two magnitudes: $\text{PE}=\text{LY}\times T_{ee}^{\text rec}$.

We will use the 2D distributions (in time and recoil energy) measured with the CsI and LAr setups~\cite{COHERENT:2020iec,COHERENT:2020ybo,COHERENT:2021xmm}. This double distribution is given by
\begin{eqnarray}
N_{ij}^{\rm signal} &=& g_j^{\rm prompt} N_i^{\rm prompt} + g_j^{\rm delayed} N_i^{\rm delayed}~,
\label{eq:NijTH}
\end{eqnarray}
where the indices $i$ and $j$ correspond to the recoil energy and time bins respectively. The $g_j^a$ factors (obtained by integrating the $g_a(t)$ functions in the $j$-th time bin) encode probability distributions for the timing of prompt/delayed events.

The relevant uncertainty sources are parametrized by suitably incorporating background events and nuisance parameters. This yields the predicted number of events as
\begin{eqnarray}
N_{ij}^{\rm th}\left( \vec{Q}_{\cal N}^2;\vec{x}\right)=N_{ij}^{\rm signal}\left( \vec{Q}_{\cal N}^2\right) \left(1 +h_{ij}^{\rm signal}(\vec{x})\right) + \sum_a N_{ij}^{\text{bkg},a}\left(1 +h_{ij}^{\text{bkg},a}(\vec{x})\right)~,
\label{eq:NiTH}
\end{eqnarray}
where $\vec{Q}_{\cal N}^2$ collects the generalized squared charges for the nucleus ${\cal N}$, $N_{ij}^{\text{bkg},a}$ are the expected number of background events of type $a$, and $h_{ij}^{\text{signal/bkg},a}(\vec{x})$ are linear functions in the nuisance parameters $\vec{x}$ that vanish at the central values $\vec{x}=\vec{0}$. 

For further details concerning the quenching factor, integration limits, efficiency, energy resolution, and $h_{ij}$ functions for the LAr and CsI analyses, we refer the reader to Ref.~\cite{Breso-Pla:2023tnz}, where our implementation of the COHERENT prescription is presented and applied to interactions involving only left-handed neutrinos.

For both LAr and CsI datasets we work with Poissonian chi-squared functions
\begin{eqnarray}
\label{eq:chi2}
\chi^2 = 
    \sum_{i, j} 2 &&\!\!\!\!\left( -N_{ij}^{\rm exp}+N_{ij}^{\rm th}\left( \vec{Q}_{\cal N}^2;\vec{x}\right) + N_{ij}^{\rm exp}~\text{ln}\left( \frac{N_{ij}^{\rm exp}}{N_{ij}^{\rm th}\left( \vec{Q}_{\cal N}^2;\vec{x}\right)}\right)\right) + \sum_n \left(\frac{x_n}{\sigma_n}\right)^2,\,~~~~~~
\end{eqnarray}
where $\sigma_n$ is the uncertainty of the nuisance parameter $x_n$. In total, we have 52x12 bins in CsI and 4x10 bins in LAr.

In addition to analyzing the existing COHERENT data, we estimate the sensitivity of the upcoming CENNS-750 detector, the next phase of COHERENT’s liquid argon CE$\nu$NS detection program~\cite{Jeong_NuFact2023,COHERENT:2022nrm}. 
For our projections, we take the SM prediction as the experimental central value. We assume that binning, detector corrections, background event counts, and source conditions remain identical to those in the current LAr measurement. Nuisance parameters are set to zero, and the number of target particles, $N_T$, is scaled to account for the projected 610 kg of active liquid argon in CENNS-750.
This serves as an illustrative study of the potential improvements achievable with increased statistics and reduced systematics. Such advancements are anticipated given the numerous ongoing and planned efforts to refine CE$\nu$NS measurements at spallation sources~\cite{Baxter:2019mcx,Barbeau:2021exu,COHERENT:2022nrm,Abele:2022iml,An:2025lws}.

\subsection{Generalized charges}\label{subsec:gen-charges}

\begin{figure}[!htb]
    \centering    
    \includegraphics[width=\linewidth]{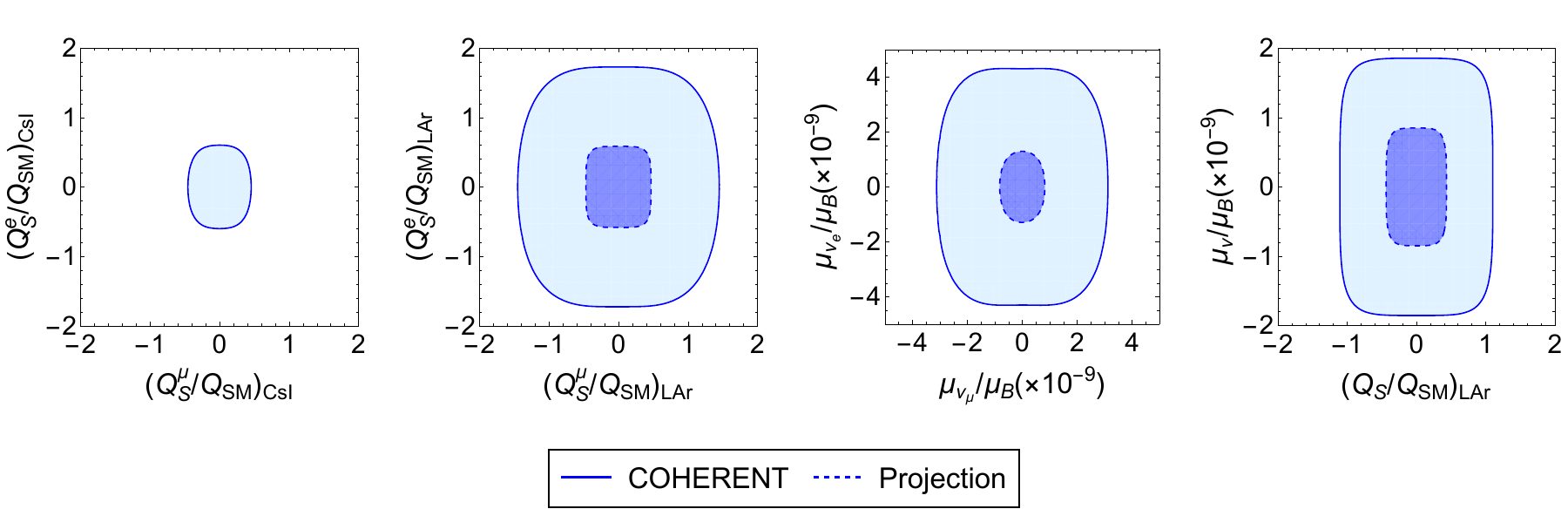}
    \hfill
    \caption{Allowed regions at 90\% CL for 
    various nonstandard scenarios. In each case, only the two parameters shown are allowed to be nonzero, in addition to the SM contribution. In the last panel, flavor universality is assumed for the effective magnetic moments and scalar charges. We consider current COHERENT data and projected measurements by the upcoming CENNS-750 detector.}
    \label{fig:F-coeffs}
\end{figure}

In this section we study some interesting scenarios, working at the level of generalized charges.
We start by assuming SM production, and work with only one free parameter, setting the rest to their SM values. Our results are shown in Table~\ref{tab:ind_charges}. 
In the case of the magnetic charges, we work directly with the effective magnetic moments, {\it cf.}~Eq.~\eqref{eq:nuclear-charges-definition}, since that allows us to combine the information from CsI and LAr measurements. Our results for $\nu_{\nu_e}$ and $\nu_{\nu_\mu}$ are similar to those found in Ref.~\cite{DeRomeri:2022twg}, which also uses the full CsI and LAr dataset.
\begin{table}
\renewcommand{\arraystretch}{1.5}
\begin{center}
\begin{tabular}{|c|c|c|}
\hline
\textsf{Generalized charge}& 
\textsf{Value (current)}& 
\textsf{Value (projected)}\\  
\hline 

$|\widetilde{Q}_V|$ (CsI) & 
$ 1.033^{+85}_{-75}\,Q_{\text{SM,CsI}} $ & -
\\

$|\widetilde{Q}_V|$ (Ar)& 
$ 1.03^{+22}_{-25}\,Q_{\text{SM,Ar}} $ & $ 1.000(12)\,Q_{\text{SM,Ar}} $
\\
\hline
\hline
\textsf{Generalized charge}& 
\textsf{90\% CL bound (current)}& 
\textsf{90\% CL bound (projected)}\\  
\hline

$|\widetilde{Q}_{S}^e|$ (CsI)  & 
$ 0.48\,Q_{\text{SM,CsI}}$ & -
\\

$|\widetilde{Q}_{S}^e|$ (Ar)& 
$ 1.5\,Q_{\text{SM,Ar}}$ & $ 0.30\,Q_{\text{SM,Ar}} $
\\

$|\widetilde{Q}_{S}^\mu|$ (CsI)& 
$ 0.39\,Q_{\text{SM,CsI}}$ & -
\\ 

$|\widetilde{Q}_{S}^\mu|$ (Ar)& 
$ 1.2\,Q_{\text{SM,Ar}}$& $ 0.19\,Q_{\text{SM,Ar}} $
\\ 

$|\mu_{\nu_e}/\mu_B|$ & 
$ 3.7 \times 10^{-9} $ & $ 1.6 \times 10^{-9} $
\\

$|\mu_{\nu_\mu}/\mu_B|$ & 
$ 2.7 \times 10^{-9} $ & $ 1.3 \times 10^{-9} $
\\
\hline
\end{tabular}
\end{center}
\caption{One-at-a-time limits on the CsI and LAr generalized nuclear charges. For the vector charges, we assume $\widetilde{Q}_V\equiv \widetilde{Q}_V^e=\widetilde{Q}_V^\mu=\widetilde{Q}_V^{\bar{\mu}}$, and we show the one-sigma results. 
For the nonstandard charges we show instead the 90\% CL bound, and we only assume $\widetilde{Q}_{S}^\mu=\widetilde{Q}_{S}^{\bar{\mu}}$.}
\label{tab:ind_charges}
 \end{table}
Fig.~\ref{fig:F-coeffs} shows the allowed regions for selected scenarios where two scalar charges $\widetilde{Q}_S^e$, $\widetilde{Q}_S^{\bar{\mu}}=\widetilde{Q}_S^\mu$ and two effective magnetic moments $\mu_{\nu_e}$, $\mu_{\nu_{\mu}}=\mu_{\nu_{\bar{\mu}}}$ are included as corrections to the SM contribution.

Let us now assume SM detection and study non-standard production effects, which are parametrized by $x_{\mu}, x_{\bar\mu}$ and $x_e$. This scenario is captured by the fit carried out in Ref.~\cite{Breso-Pla:2023tnz} using flavor-dependent vector charges, which gives 
\begin{eqnarray}
    \begin{pmatrix}
        x_{\mu} \\ x_{\bar{\mu}} \\ x_e
    \end{pmatrix} =
    \begin{pmatrix}
        1.30(32)  \\ -1.2(1.4) \\ 4.3(2.2) 
    \end{pmatrix},~\rho = 
    \begin{pmatrix}
         1 & 0.16 & -0.14  \\ \,\, - & 1 & -0.98 \\ - & - & 1 
    \end{pmatrix}~.
\end{eqnarray}
These results can be rewritten as the following uncorrelated bounds
\begin{eqnarray}    
    \begin{pmatrix}
        0.15\, x_{\mu} - 0.83 \, x_{\bar{\mu}} - 0.53\, x_e \\
        0.99 \, x_{\mu} + 0.12 \, x_{\bar{\mu}} + 0.10\, x_e \\
        -0.02\, x_{\mu} - 0.54 \, x_{\bar{\mu}} + 0.84\, x_e 
    \end{pmatrix} =
    \begin{pmatrix}
         -1.05(22) \\ 1.55(32) \\ 4.2(2.6)
    \end{pmatrix}~.
\end{eqnarray}
Allowing for non-standard effects only in pion decay (i.e., $x_{\bar\mu}=x_e=1$), we find
    \begin{equation}
    x_{\mu} = 1.41 \pm 0.32~\text{(current)}~\to 1.000\pm 0.084~\text{(projection)}~. 
    \end{equation}
Allowing instead for non-standard effects only in muon decay (i.e., $x_\mu =1$), we find
\begin{equation}
    \left( \begin{matrix}x_{\bar{\mu}} \\ x_e
    \end{matrix} \right) 
    = \left( \begin{matrix} \,\, -1.5 \pm 1.3 \\ 4.6 \pm 2.1 
    \end{matrix} \right), 
    ~ \quad \rho = -0.98~,
\end{equation}
which can be rewritten as the following uncorrelated bounds
\begin{eqnarray}
\begin{matrix}
0.84\, x_{\bar{\mu}} + 0.54\, x_e \\
0.54\, x_{\bar{\mu}} - 0.84\,x_e 
\end{matrix} && 
\left.\begin{matrix}
    \hspace{-0.2cm}=\,\, 1.25 \pm 0.21 \\ 
    =\,\, -4.70 \pm 2.5
\end{matrix}\right\}~~ \text{(current)} \nn
\begin{matrix}
0.81\, x_{\bar{\mu}} + 0.58\, x_e \\
0.58\, x_{\bar{\mu}} - 0.81\,x_e 
\end{matrix} && 
\left.\begin{matrix}
    \hspace{-0.2cm} =\,\, 1.39 \pm 0.04 \\ 
    = -0.23 \pm 0.49
\end{matrix}\right\}~~ \text{(projection)}
\end{eqnarray}

Let us rewrite the first bound in terms of the $P_x$ and $w_x$ parameters
\begin{eqnarray}\label{eq:P-w-bound}
    0.54\,P_{\nu_L} + 0.84\,P_{\bar\nu_L} + 0.4\,\left(P_{\nu_L}w_{\nu_L} - P_{\bar\nu_L}w_{\bar\nu_L}\right) = 1.25 \pm 0.21~.
\end{eqnarray}
This result was presented in our recent letter~\cite{Breso-Pla:2025cul} for the simpler case of lepton-flavor conserving muon decay with three right-handed neutrinos. The general calculation carried out in this work shows that the result is parametrically the same for an arbitrary number of right-handed neutrinos when LFV is allowed, in other words, when all (anti)neutrino flavors are present in the decay of the muon. This constrains, e.g., interactions involving tau neutrinos, as discussed in the next subsection.

Our results provide the first probe of antineutrino parameters $P_{\bar\nu_L}$ and $w_{\bar\nu_L}$~\cite{Breso-Pla:2025cul}. They also constitute the second study of the neutrino parameters $P_{\nu_L}$ and $w_{\nu_L}$ following the KARMEN experiment~\cite{PhysRevLett.81.520}. The latter was based on W. Fetscher's theoretical work~\cite{Fetscher:1994nv} and used the charged-current reaction ${}^{12}C(\nu_e, e^-){}^{12}N_{\text{g.s.}}$ to measure the energy distribution of $\nu_e$ emitted in muon decay.

\subsection{$\nu$WEFT Wilson Coefficients}\label{subsec:nuWEFt-WCs}
In this section we move up in the theory description and focus on the underlying $\nu$WEFT Wilson coefficients. The one-at-a-time bounds that we obtain for coefficients relevant for detection (CE$\nu$NS) and production through pion decay are shown in Table~\ref{tab:vWEFT-bounds}. The NC vector coefficients with right-handed neutrinos, like $\tilde\epsilon_{ee}^{uu}$, are not shown in the table because they affect the observable only if nonstandard production is present. Simply put, a right-handed neutrino can only be detected if it is produced at the source. On the other hand, the semileptonic CC axial-vector coefficients with right-handed neutrinos, $\tilde\epsilon_L^{ud}-\tilde\epsilon_R^{ud}$, can be probed without invoking nonstandard physics in detection due to their indirect effect in the pion decay width, namely, the denominator of $x_\mu$ in Eq.~\eqref{eq:xdefinitions}. 

\begin{table}[bht]
    \centering
    \renewcommand{\arraystretch}{1.5}
    \begin{tabular}{|c|c|c|c|c|c|}
         \hline
         \multirow{2}{*}{\textsf{Coefficient}}&
         \multicolumn{2}{c|}{\textsf{Bound (90\% C.L.)}}&
         \multirow{2}{*}{\textsf{Coefficient}}&
         \multicolumn{2}{c|}{\textsf{Bound (90\% C.L.)}}\\
         \cline{2-3}\cline{5-6}
         
         &
         \textsf{Current}&
         \textsf{Projected}&
         &
         \textsf{Current}&
         \textsf{Projected}\\
         \hline
         \hline
         \multicolumn{3}{|c|}{(I)}&  
         \multicolumn{3}{c|}{(II)}\\
         \hline

         $\left|\epsilon_{ee}^{uu}\right|$&
         0.078 &  0.035
         &
         $|\epsilon_{e\mu}^{uu}|$&
         0.12& 0.016
         \\

         $\left|\epsilon_{\mu\mu}^{uu}\right|$&
         0.049 & 0.016
         & 
         $|\epsilon_{e\tau}^{uu}|$&
         0.17& 0.030
         \\

         $\left|\epsilon_{ee}^{dd}\right|$&
         0.071 & 0.033
         & 
         $|\epsilon_{\mu\tau}^{uu}|$&
         0.15& 0.019
         \\

         $\left|\epsilon_{\mu\mu}^{dd}\right|$&
         0.043 & 0.015
         &
         $|\epsilon_{e\mu}^{dd}|$&
         0.11& 0.015
         \\
         \cline{1-3}

         \multicolumn{3}{|c|}{(III)}&
         $|\epsilon_{e\tau}^{dd}|$&
         0.15& 0.028
         \\
         \cline{1-3}

         $|[\tilde\epsilon_S^{uu}]_{e\alpha}|$&
         $1.6 \times 10^{-2}$& $9.1 \times 10^{-3}$
         &
         $|\epsilon_{\mu\tau}^{dd}|$&
         0.13& 0.018
         \\
         \cline{4-6}

         $|[\tilde\epsilon_S^{uu}]_{\mu\alpha}|$&
         $1.4 \times 10^{-2}$& $5.8 \times 10^{-3}$
         &
         \multicolumn{3}{c|}{(IV)}\\
         \cline{4-6}

         $|[\tilde\epsilon_S^{dd}]_{e\alpha}|$&
         $1.5 \times 10^{-2}$ & $8.9 \times 10^{-3}$
         & 
         $|[\tilde\epsilon_L^{ud}-\tilde\epsilon_R^{ud}]_{\mu\alpha}|$&
         $0.51$ & 0.30
         \\

         $|[\tilde\epsilon_S^{dd}]_{\mu\alpha}|$&
         $1.3 \times 10^{-2}$& $5.7 \times 10^{-3}$
         &
         $|[\tilde\epsilon^{ud}_P]_{\mu\alpha}|$&
         $0.019$& 0.011
         \\
         \hline

    \end{tabular}
    \caption{One-at-a-time bounds on $\nu$WEFT Wilson coefficients belonging to the categories (I) Flavour diagonal NC vector WCs (II) Flavour off-diagonal NC vector WCs (III) Scalar NC WCs and (IV) Semi-leptonic CC WCs. The flavor subindex $\alpha$ can have any value, i.e., $\alpha=e,\mu,\tau$. We omit the so-called \textit{dark solutions} that appear for large $\epsilon^{qq}_{\alpha\alpha}$ values away from the SM.}
    \label{tab:vWEFT-bounds}
\end{table}

For operators affecting muon decay the discussion is more complex. 
First, let us note that the combination ${\rm Tr}\left[|h^V_{LL}|^2+\frac{1}{4}|h^S_{RR}|^2\right]$ appears together in both $x_{\bar\mu}$ and $x_e$, and thus cannot be disentangled using only COHERENT data. Setting to zero other operators affecting the event rate (i.e., $h^V_{LR}=h^V_{RL}=h^S_{LR}=h^S_{RL}=h^T_{LR}=h^T_{RL}=0$) we obtain the following one-$\sigma$ bound
\begin{align}\label{eq:flat}
    {\rm Tr}\left[|h^V_{LL}|^2+ \frac{1}{4}|h^S_{RR}|^2\right] = 0.95^{+0.05}_{-0.15}~,
\end{align}
in agreement with one, the SM result. Equivalently, we find a 90\% CL lower bound of 0.72. Assuming $h^S_{RR}=0$ we obtain the bound on $[h^V_{LL}]_{\alpha\beta}$ shown in the first line of Table~\ref{tab:muon-decay-params-bounds}.

Using the normalization condition, Eq.~(\ref{eq:muon_decay_norm}), this result can be expressed as an indirect constraint on the remaining Wilson coefficients (which do not contribute to COHERENT in the case of SM detection), $h^V_{RR}$ and $h^S_{LL}$, namely
\begin{align}
    {\rm Tr}\left[|h^V_{RR}|^2+ \frac{1}{4}|h^S_{LL}|^2\right] &
    < 0.28~(90~\text{\% CL)}~, 
\end{align}
which leads to the bounds shown in the last two rows of Table~\ref{tab:muon-decay-params-bounds}. We note that COHERENT is not able to disentangle the various flavor structures $[h^X_{\epsilon\eta}]_{\alpha\beta}$ because it is only sensitive to the quadratic combination ${\rm Tr}\left[|h^X_{\epsilon\eta}|^2\right]=\sum_{\alpha\beta} |[h^X_{\epsilon\eta}]_{\alpha\beta}|^2$. This is due to the flavor independence of the CE$\nu$NS cross section in the SM.

\begin{figure}[t]
    \centering
    \includegraphics[width=\linewidth]{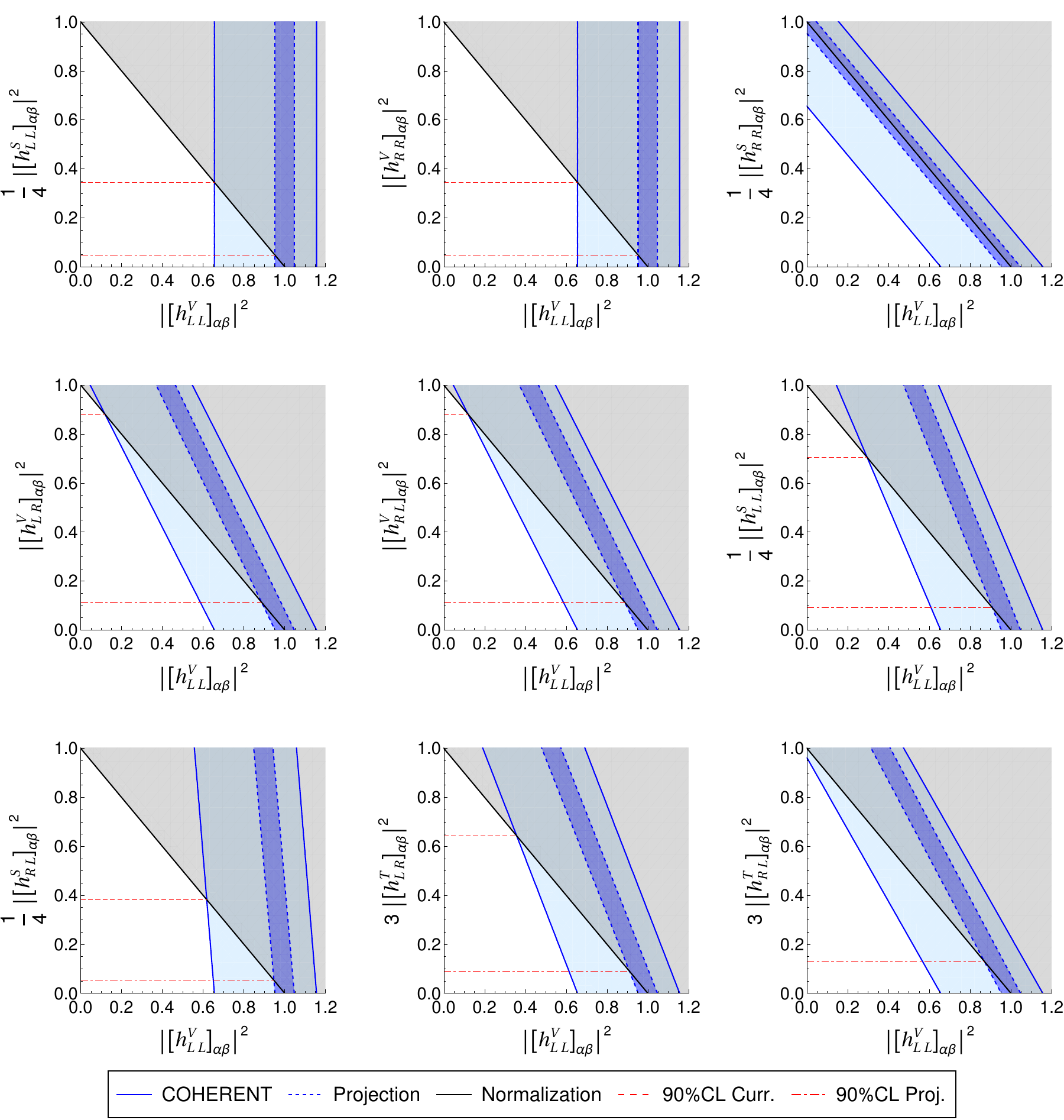}
    \caption{Allowed regions for $\nu$WEFT Wilson coefficients that parametrize muon decay. Only the two parameters shown in each panel are non-zero. The black diagonal solid line indicates the normalization condition, which the two couplings must satisfy. The light (dark) blue area indicates the $\Delta\chi^2=2.71$ region obtained in current (future) COHERENT data, so that its intersection with the normalization line gives the 90\% CL region. 
    The red horizontal lines indicate the corresponding limits on the nonstandard coefficient displayed in the vertical axes.\vspace{1cm}}
     \label{fig:h-X-ab-2Dplots}
\end{figure}

\begin{table}[t]
    \centering
    \renewcommand{\arraystretch}{1.6}
    \begin{tabular}{|c|c|c|c|}
    \hline
    \multirow{2}{*}{\textsf{Coefficient}}&
    \multicolumn{3}{c|}{\textsf{Bound (90\% C.L.)}}\\
    \cline{2-4}
    
    &
    \textsf{Current}&
    \textsf{Projected}&
    \textsf{PDG \small{$(\alpha = \mu, \beta = e)$}}\\
    \hline\hline

    $|[h^V_{LL}]_{\alpha\beta}|$&
    >0.848&
    >0.976&
    $>$0.960\\ \hline

    $|[h^V_{LR}]_{\alpha\beta}|$& 0.79
    & 0.34
    &
    0.023\\

    $|[h^V_{RL}]_{\alpha\beta}|$& 0.79
    & 0.34
    &
    0.105\\

    $|[h^S_{LR}]_{\alpha\beta}|$& 1.4
    & 0.57
    &
    0.050\\

    $|[h^S_{RL}]_{\alpha\beta}|$& 1.6
    & 0.67
    &
    0.420\\

    $|[h^T_{LR}]_{\alpha\beta}|$& 0.47
    & 0.21
    &
    0.015\\

    $|[h^T_{RL}]_{\alpha\beta}|$& 0.42
    & 0.17
    &
    0.105\\

    \hline
    $|[h^V_{RR}]_{\alpha\beta}|~(*)$&
    0.53&
    0.22&
    0.017\\

    $|[h^S_{LL}]_{\alpha\beta}|~(*)$&
    1.07&
    0.44&
    0.550\\

    \hline
    \end{tabular}
    \caption{Bounds on $\nu$WEFT Wilson coefficients that parametrize muon decay, assuming SM for pion decay and CE$\nu$NS. For the first block, i.e. $h^V_{LL}$, the {\it lower} bound is obtained assuming only that coefficient contributes to COHERENT ($h^V_{RR}$ and $h^S_{LL}$ are also non-zero but they do not contribute to the event rate). 
    For each line of the second block we carry out a two parameter fit with $h^V_{LL}$ and the displayed coefficient, taking into account they must satisfy the normalization condition (see Fig.~\ref{fig:h-X-ab-2Dplots}). 
    For the coefficients in the third block, we carry out again a two parameter fit with $h^V_{LL}$ and the displayed coefficient. Although the latter does not contribute to COHERENT, the indirect limits shown in the table, highlighted with $(*)$, are obtained from the $h^V_{LL}$ bound by employing the normalization condition, {\it cf.} Eq.~\eqref{eq:flat}.  
    We also present the projected improvements based on future CENNS-750 measurements and compare with the limits reported by the PDG~\cite{MuonDecayParameters:2024cfk}.
    The flavor subindices $\alpha$ and $\beta$ can have any value, i.e., $\alpha,\beta=e,\mu,\tau$.}
    \label{tab:muon-decay-params-bounds}
\end{table}

Let us discuss now the other Wilson Coefficients contributing to muon decay. We cannot do one-parameter fits because if only one Wilson Coefficient is present, the normalization condition fixes its exact value. The simplest meaningful fits have two parameters, $[h^V_{LL}]_{\mu e}$ and an exotic one, and are studied in Fig.~\ref{fig:h-X-ab-2Dplots}. The normalization condition forces the two couplings shown in each panel to stay inside the black diagonal line\footnote{The region below the black diagonal line becomes physical if one switches on  the $h^V_{RR}$ and $h^S_{LL}$ coefficients, which do not contribute to COHERENT.}, so we are effectively working with one degree of freedom. Thus, we use $\Delta\chi^2=2.71$ for the 90\% CL regions, in such a way that the red horizontal lines indicate the corresponding limits on the vertical axes. The results can also be found in numerical form in Table~\ref{tab:muon-decay-params-bounds}, along with estimated projections. 

The scenario in the upper left corner, where only $h^V_{LL}$ and $h^S_{LL}$ are present, is particularly interesting because traditional muon decay measurements (where only the electron/positron is detected) cannot provide any constrain. They can only determine the combination ${\rm Tr}\left[|h^V_{LL}|^2+\frac{1}{4}|h^S_{LL}|^2\right]$, which is already fixed to one by the normalization condition, Eq.~\eqref{eq:muon_decay_norm}. 
Past works focused on the lepton-flavor conserving case ($\mu^+\to e^+\nu_e\bar\nu_\mu$) and broke this flat direction using inverse muon decay data ($\nu_\mu e^-\to \mu^- \nu_e$)~\cite{MISHRA1990170,CHARM-II:1995xfh}, which is governed by the same interactions. Our results show that COHERENT can also disentangle these two couplings. This was used for the first time in our recent letter~\cite{Breso-Pla:2025cul}, and the result was compared with the bounds from inverse muon decay and KARMEN~\cite{PhysRevLett.81.520}. The latter is the only other work that uses the detection of $\nu_e$ emitted in muon decay to probe muon decay parameters. We show in Ref.~\cite{Breso-Pla:2025cul} that the current bound from COHERENT data is not competitive, but future CE$\nu$NS data have the potential to provide the best bound. 

For simplicity we do not study more complex scenarios where more than 2 parameters contribute to COHERENT observables. In that case, additional flat directions will appear because CE$\nu$NS data can only extract two specific combinations, namely, $x_{\bar\mu}$ and $x_e$.

\section{Conclusions}\label{sec:conclusions}

In this paper, we have outlined a meticulous investigation of neutrino-nucleus scattering data from the COHERENT experiment in the backdrop of the $\nu$WEFT, the low-energy EFT that includes light SM fields along with right-handed Dirac neutrino fields. We studied new physics effects in full generality, by taking into account, for the first time, flavor-general Wilson coefficients affecting both (anti)neutrino production via pion/muon decay, and detection via the CE$\nu$NS process. For the latter, we used a series of EFTs to go from the underlying interactions involving quarks, to the scattering amplitudes involving nuclear targets, through the intermediate scale of nucleon-lepton interactions.

Despite the generality of our approach, the event rate can be written in a compact form as a convolution of SM fluxes and effective cross-sections, where all non-standard effects are encapsulated by a few effective nuclear charges, {\it cf.}~Eqs.~\eqref{eq:prompt_and_delayedBIS}-\eqref{eq:Qsq-full-exprs}. This facilitates the implementation of this general approach in existing or future analyses of CE$\nu$NS data.

We studied interesting limits of our general expressions, and recovered results previously obtained in the literature for simplified cases. We showed, within the non-relativistic nucleon EFT, that the tensor contribution does not enjoy the coherent enhancement (at zeroth order in recoil), in agreement with Refs.~\cite{Altmannshofer:2018xyo,Hoferichter:2020osn,Glick-Magid:2023uhk}. This observation has been overlooked in other works.  

We demonstrated how the study of the COHERENT experiment, through the lens of the $\nu$WEFT,
provides novel access to the physics of neutrino production (muon and pion decay). This work generalizes the findings of our recent letter~\cite{Breso-Pla:2025cul}, which assumed SM detection and lepton-flavor conservation.

We conducted a statistical analysis of available COHERENT data, based on scattering against LAr and CsI targets~\cite{COHERENT:2020iec,COHERENT:2020ybo,COHERENT:2021xmm} to extract a variety of quantities. We extracted the vector charge, assuming lepton flavor universality, and set bounds on the $\nu_e$ and $\nu_\mu$ scalar charges and effective magnetic moments, for one- and two-at-a-time scenarios. For each case we also estimated projected improvements based on the proposed 610 kg LAr based detector, CENNS-750 \cite{Jeong_NuFact2023,COHERENT:2022nrm}. Thus, we illustrated the potential gains from higher statistics and lower systematics, expected from the ongoing and planned efforts to refine CE$\nu$NS measurements~\cite{Baxter:2019mcx,Barbeau:2021exu,COHERENT:2022nrm,Abele:2022iml,An:2025lws}.
We also obtained bounds on the $x_f$ production factors, for cases with new physics (i) only in pion decay, (ii) only in muon decay and (iii) in both production channels. We extracted the implications for the muon-decay parameters describing the neutrino and antineutrino energy distributions, generalizing the results recently obtained in Ref.~\cite{Breso-Pla:2025cul} to a generic scenario with an arbitrary number of right-handed neutrinos and without any restrictions on lepton-flavor violation. This substantiates the wide applicability of our $\nu$WEFT analysis. Finally, we utilized the results obtained for generalized charges and production factors to infer limits (and projections) on $\nu$WEFT Wilson coefficients with an arbitrary flavor structure. These include NC operators affecting CE$\nu$NS, semileptonic CC operators impacting pion decay, and 4-lepton CC operators relevant for muon decay. These results can be straightforwardly applied to specific NP models or underlying EFTs. For instance, translating these bounds to the $\nu$SMEFT~\cite{Chala:2020vqp} enables the comparison of CE$\nu$NS measurements at spallation sources with collider searches, electroweak precision observables, and LFV probes within those scenarios.

We conclude by noting that our approach can also be extended to recent CE$\nu$NS measurements with reactor and solar neutrinos~\cite{Colaresi:2022obx,XENON:2024ijk,PandaX:2024muv,Ackermann:2025obx}, to include not only nonstandard NC effects in detection (and propagation), but also CC effects in the corresponding production processes, such as nuclear beta decays.

\acknowledgments
We thank Adam Falkowski and Martin Novoa-Brunet for useful discussions.
This work has been supported by MCIN/AEI/10.13039/501100011033 (grants PID2020-114473GB-I00 and PID2023-146220NB-I00) and by MICIU/AEI/10.13039/501100011033 and European Union NextGenerationEU/PRTR (grant CNS2022-135595). VB is supported by the DFG under grant
396021762 – TRR 257: \textit{Particle Physics Phenomenology after the Higgs Discovery} and the BMBF Junior Group \textit{Generative Precision Networks for Particle Physics} (DLR 01IS22079).

\appendix
\section{Kinematic $f_X^j(T)$ functions}
\label{app:fullresults}
The kinematical functions for prompt events (pion decay) are given by
\begin{align}
    f_{_V}^{\mu}(T) &= 1-\frac{m_\mathcal{N}T}{2E^2_{\nu,\pi}}-\frac{T}{E_{\nu,\pi}}, \qquad\qquad
    f_{_S}^\mu(T) = \frac{m_\mathcal{N} T}{2E^2_{\nu,\pi}}, \nn
    f_{_F}^{\mu}(T) &= \frac{v^2}{2m_\mathcal{N} T}\left(1-\frac{T}{E_{\nu,\pi}}\right), \qquad\quad
    f_{_{SF}}^{\mu}(T) = \frac{v}{2E_{\nu,\pi}}\left(1-\frac{T}{2E_{\nu,\pi}}\right),
\end{align}
where $E_{\nu,\pi}=(m_{\pi^\pm}^2-m_\mu^2)/(2m_{\pi^\pm})$ is the fixed energy of the neutrino in the decay. 
The kinematic functions for delayed events (muon decay) are

\begin{eqnarray}
        f_{_V}^{\bar{\mu}}(T) &=& 16\left[-\frac{3m_{\mathcal{N}}T}{2m^2_\mu}\frac{E_\nu}{m_\mu}-\frac{T}{m_\mu}\left(\frac{3}{2}-\frac{m_\mathcal{N}}{m_\mu}\right)\frac{E_\nu^2}{m_\mu^2}+\left(1+\frac{4T}{3m_\mu}\right)\frac{E_\nu^3}{m_\mu^3}-\frac{E_\nu^4}{m_\mu^4}\right]_{E_\nu^{min}}^{m_\mu/2}~,
        \nn
        f_{_S}^{\bar{\mu}}(T) &=& 16\,\frac{m_\mathcal{N}T}{m_\mu^2}\left[\frac{3}{2}\frac{E_\nu}{m_\mu}-\frac{E_\nu^2}{m_\mu^2}\right]_{E_\nu^{min}}^{m_\mu/2}~,\nn
        f_{_F}^{\bar{\mu}}(T) &=& \frac{8v^2}{m_\mathcal{N}T}\left[-\frac{3T}{2m_\mu}\frac{E_\nu^2}{m_\mu^2}+\left(1+\frac{4T}{3m_\mu}\right)\frac{E_\nu^3}{m_\mu^3}-\frac{E_\nu^4}{m_\mu^4}\right]_{E_\nu^{min}}^{m_\mu/2}~,
        \nn
        f_{_{SF}}^{\bar{\mu}}(T) &=& \frac{8v}{m_\mu}\left[-\frac{3T}{2m_\mu}\frac{E_\nu}{m_\mu}+\left(\frac{3}{2}+\frac{T}{m_\mu}\right)\frac{E_\nu^2}{m_\mu^2}-\frac{4}{3}\frac{E_\nu^3}{m_\mu^3}\right]_{E_\nu^{min}}^{m_\mu/2}~,
        \nn
        f_{_V}^{e}(T) &=& 16\left[-\frac{3m_{\mathcal{N}}T}{m^2_\mu}\frac{E_\nu}{m_\mu}-\frac{3T}{m_\mu}\left(1-\frac{m_\mathcal{N}}{m_\mu}\right)\frac{E_\nu^2}{m_\mu^2}+\left(2+\frac{4T}{m_\mu}\right)\frac{E_\nu^3}{m_\mu^3}-3\frac{E_\nu^4}{m_\mu^4}\right]_{E_\nu^{min}}^{m_\mu/2}~, \nn
        f_{_S}^e(T) &=& 48\,\frac{m_\mathcal{N}T}{m_\mu^2}\left[\frac{E_\nu}{m_\mu}-\frac{E_\nu^2}{m_\mu^2}\right]_{E_\nu^{min}}^{m_\mu/2}~,
        \nn
        f_{_F}^{e}(T) &=& \frac{8v^2}{m_\mathcal{N}T}\left[-3\frac{T}{m_\mu}\frac{E_\nu^2}{m_\mu^2}+\left(2+\frac{4T}{m_\mu}\right)\frac{E_\nu^3}{m_\mu^3}-3\frac{E_\nu^4}{m_\mu^4}\right]_{E_\nu^{min}}^{m_\mu/2}~,
        \nn
        f_{_{SF}}^{e}(T) &=& \frac{24v}{m_\mu}\left[-\frac{T}{m_\mu}\frac{E_\nu}{m_\mu}+\left(1+\frac{T}{m_\mu}\right)\frac{E_\nu^2}{m_\mu^2}-\frac{4}{3}\frac{E_\nu^3}{m_\mu^3}\right]_{E_\nu^{min}}^{m_\mu/2}~,        
\end{eqnarray}
where $E_\nu^{min}=\left(T/2\right)\left(1+\sqrt{1+2\,m_\mathcal{N}/T}\right)$.

{
\bibliographystyle{JHEP}
\small
\bibliography{COHERENT}

\providecommand{\href}[2]{#2}\begingroup\raggedright\begin{thebibliography}{10}

\bibitem{Freedman:1973yd}
D.Z.~Freedman, {{Coherent Neutrino Nucleus Scattering as a Probe of the Weak
  Neutral Current}}, \href{https://doi.org/10.1103/PhysRevD.9.1389}{\emph{Phys.
  Rev. D} {\bfseries 9} (1974) 1389}.

\bibitem{COHERENT:2017ipa}
{\scshape COHERENT} collaboration, {{Observation of Coherent Elastic
  Neutrino-Nucleus Scattering}},
  \href{https://doi.org/10.1126/science.aao0990}{\emph{Science} {\bfseries 357}
  (2017) 1123} [\href{https://arxiv.org/abs/1708.01294}{{\ttfamily
  1708.01294}}].

\bibitem{COHERENT:2020iec}
{\scshape COHERENT} collaboration, {{First Measurement of Coherent Elastic
  Neutrino-Nucleus Scattering on Argon}},
  \href{https://doi.org/10.1103/PhysRevLett.126.012002}{\emph{Phys. Rev. Lett.}
  {\bfseries 126} (2021) 012002}
  [\href{https://arxiv.org/abs/2003.10630}{{\ttfamily 2003.10630}}].

\bibitem{COHERENT:2020ybo}
{\scshape COHERENT} collaboration, {{COHERENT Collaboration data release from
  the first detection of coherent elastic neutrino-nucleus scattering on
  argon}},  \href{https://arxiv.org/abs/2006.12659}{{\ttfamily 2006.12659}}.

\bibitem{COHERENT:2021xmm}
{\scshape COHERENT} collaboration, {{Measurement of the Coherent Elastic
  Neutrino-Nucleus Scattering Cross Section on CsI by COHERENT}},
  \href{https://doi.org/10.1103/PhysRevLett.129.081801}{\emph{Phys. Rev. Lett.}
  {\bfseries 129} (2022) 081801}
  [\href{https://arxiv.org/abs/2110.07730}{{\ttfamily 2110.07730}}].

\bibitem{Colaresi:2022obx}
J.~Colaresi, J.I.~Collar, T.W.~Hossbach, C.M.~Lewis and K.M.~Yocum,
  {{Measurement of Coherent Elastic Neutrino-Nucleus Scattering from Reactor
  Antineutrinos}},
  \href{https://doi.org/10.1103/PhysRevLett.129.211802}{\emph{Phys. Rev. Lett.}
  {\bfseries 129} (2022) 211802}
  [\href{https://arxiv.org/abs/2202.09672}{{\ttfamily 2202.09672}}].

\bibitem{Ackermann:2025obx}
N.~Ackermann et~al., {{First observation of reactor antineutrinos by coherent
  scattering}},  \href{https://arxiv.org/abs/2501.05206}{{\ttfamily
  2501.05206}}.

\bibitem{CCM:2021leg}
{\scshape CCM} collaboration, {{First dark matter search results from Coherent
  CAPTAIN-Mills}},
  \href{https://doi.org/10.1103/PhysRevD.106.012001}{\emph{Phys. Rev. D}
  {\bfseries 106} (2022) 012001}
  [\href{https://arxiv.org/abs/2105.14020}{{\ttfamily 2105.14020}}].

\bibitem{Zema:2020qen}
{\scshape COSINUS} collaboration, {{COSINUS: A NaI-based cryogenic calorimeter
  for direct dark matter search}},
  \href{https://doi.org/10.1393/ncc/i2019-19228-1}{\emph{Nuovo Cim. C}
  {\bfseries 42} (2020) 228}.

\bibitem{CONNIE:2024pwt}
{\scshape CONNIE} collaboration, {{Searches for CE\ensuremath{\nu}NS and
  Physics beyond the Standard Model using Skipper-CCDs at CONNIE}},
  \href{https://arxiv.org/abs/2403.15976}{{\ttfamily 2403.15976}}.

\bibitem{CONUS:2020skt}
{\scshape CONUS} collaboration, {{Constraints on elastic neutrino nucleus
  scattering in the fully coherent regime from the CONUS experiment}},
  \href{https://doi.org/10.1103/PhysRevLett.126.041804}{\emph{Phys. Rev. Lett.}
  {\bfseries 126} (2021) 041804}
  [\href{https://arxiv.org/abs/2011.00210}{{\ttfamily 2011.00210}}].

\bibitem{XENON:2024ijk}
{\scshape XENON} collaboration, {{First Indication of Solar B8 Neutrinos via
  Coherent Elastic Neutrino-Nucleus Scattering with XENONnT}},
  \href{https://doi.org/10.1103/PhysRevLett.133.191002}{\emph{Phys. Rev. Lett.}
  {\bfseries 133} (2024) 191002}
  [\href{https://arxiv.org/abs/2408.02877}{{\ttfamily 2408.02877}}].

\bibitem{PandaX:2024muv}
{\scshape PandaX} collaboration, {{First Indication of Solar B8 Neutrinos
  through Coherent Elastic Neutrino-Nucleus Scattering in PandaX-4T}},
  \href{https://doi.org/10.1103/PhysRevLett.133.191001}{\emph{Phys. Rev. Lett.}
  {\bfseries 133} (2024) 191001}
  [\href{https://arxiv.org/abs/2407.10892}{{\ttfamily 2407.10892}}].

\bibitem{Barranco:2005yy}
J.~Barranco, O.~Miranda and T.I.~Rashba, {{Probing new physics with coherent
  neutrino scattering off nuclei}},
  \href{https://doi.org/10.1088/1126-6708/2005/12/021}{\emph{JHEP} {\bfseries
  12} (2005) 021} [\href{https://arxiv.org/abs/hep-ph/0508299}{{\ttfamily
  hep-ph/0508299}}].

\bibitem{Scholberg:2005qs}
K.~Scholberg, {{Prospects for measuring coherent neutrino-nucleus elastic
  scattering at a stopped-pion neutrino source}},
  \href{https://doi.org/10.1103/PhysRevD.73.033005}{\emph{Phys. Rev. D}
  {\bfseries 73} (2006) 033005}
  [\href{https://arxiv.org/abs/hep-ex/0511042}{{\ttfamily hep-ex/0511042}}].

\bibitem{Cadeddu:2017etk}
M.~Cadeddu, C.~Giunti, Y.F.~Li and Y.Y.~Zhang, {{Average CsI neutron density
  distribution from COHERENT data}},
  \href{https://doi.org/10.1103/PhysRevLett.120.072501}{\emph{Phys. Rev. Lett.}
  {\bfseries 120} (2018) 072501}
  [\href{https://arxiv.org/abs/1710.02730}{{\ttfamily 1710.02730}}].

\bibitem{Papoulias:2017qdn}
D.K.~Papoulias and T.S.~Kosmas, {{COHERENT constraints to conventional and
  exotic neutrino physics}},
  \href{https://doi.org/10.1103/PhysRevD.97.033003}{\emph{Phys. Rev. D}
  {\bfseries 97} (2018) 033003}
  [\href{https://arxiv.org/abs/1711.09773}{{\ttfamily 1711.09773}}].

\bibitem{Shoemaker:2017lzs}
I.M.~Shoemaker, {{COHERENT search strategy for beyond standard model neutrino
  interactions}}, \href{https://doi.org/10.1103/PhysRevD.95.115028}{\emph{Phys.
  Rev. D} {\bfseries 95} (2017) 115028}
  [\href{https://arxiv.org/abs/1703.05774}{{\ttfamily 1703.05774}}].

\bibitem{Liao:2017uzy}
J.~Liao and D.~Marfatia, {{COHERENT constraints on nonstandard neutrino
  interactions}},
  \href{https://doi.org/10.1016/j.physletb.2017.10.046}{\emph{Phys. Lett. B}
  {\bfseries 775} (2017) 54}
  [\href{https://arxiv.org/abs/1708.04255}{{\ttfamily 1708.04255}}].

\bibitem{Cadeddu:2018dux}
M.~Cadeddu, C.~Giunti, K.A.~Kouzakov, Y.-F.~Li, Y.-Y.~Zhang and
  A.I.~Studenikin, {{Neutrino Charge Radii From Coherent Elastic
  Neutrino-nucleus Scattering}},
  \href{https://doi.org/10.1142/9789811233913_0013}{\emph{Phys. Rev. D}
  {\bfseries 98} (2018) 113010}
  [\href{https://arxiv.org/abs/1810.05606}{{\ttfamily 1810.05606}}].

\bibitem{AristizabalSierra:2018eqm}
D.~Aristizabal~Sierra, V.~De~Romeri and N.~Rojas, {{COHERENT analysis of
  neutrino generalized interactions}},
  \href{https://doi.org/10.1103/PhysRevD.98.075018}{\emph{Phys. Rev. D}
  {\bfseries 98} (2018) 075018}
  [\href{https://arxiv.org/abs/1806.07424}{{\ttfamily 1806.07424}}].

\bibitem{Denton:2018xmq}
P.B.~Denton, Y.~Farzan and I.M.~Shoemaker, {{Testing large non-standard
  neutrino interactions with arbitrary mediator mass after COHERENT data}},
  \href{https://doi.org/10.1007/JHEP07(2018)037}{\emph{JHEP} {\bfseries 07}
  (2018) 037} [\href{https://arxiv.org/abs/1804.03660}{{\ttfamily
  1804.03660}}].

\bibitem{Altmannshofer:2018xyo}
W.~Altmannshofer, M.~Tammaro and J.~Zupan, {{Non-standard neutrino interactions
  and low energy experiments}},
  \href{https://doi.org/10.1007/JHEP11(2021)113}{\emph{JHEP} {\bfseries 09}
  (2019) 083} [\href{https://arxiv.org/abs/1812.02778}{{\ttfamily
  1812.02778}}].

\bibitem{Giunti:2019xpr}
C.~Giunti, {{General COHERENT constraints on neutrino nonstandard
  interactions}},
  \href{https://doi.org/10.1103/PhysRevD.101.035039}{\emph{Phys. Rev. D}
  {\bfseries 101} (2020) 035039}
  [\href{https://arxiv.org/abs/1909.00466}{{\ttfamily 1909.00466}}].

\bibitem{Coloma:2019mbs}
P.~Coloma, I.~Esteban, M.C.~Gonzalez-Garcia and M.~Maltoni, {{Improved global
  fit to Non-Standard neutrino Interactions using COHERENT energy and timing
  data}}, \href{https://doi.org/10.1007/JHEP02(2020)023}{\emph{JHEP} {\bfseries
  02} (2020) 023} [\href{https://arxiv.org/abs/1911.09109}{{\ttfamily
  1911.09109}}].

\bibitem{Skiba:2020msb}
W.~Skiba and Q.~Xia, {{Electroweak constraints from the COHERENT experiment}},
  \href{https://doi.org/10.1007/JHEP10(2022)102}{\emph{JHEP} {\bfseries 10}
  (2022) 102} [\href{https://arxiv.org/abs/2007.15688}{{\ttfamily
  2007.15688}}].

\bibitem{Hoferichter:2020osn}
M.~Hoferichter, J.~Men\'endez and A.~Schwenk, {{Coherent elastic
  neutrino-nucleus scattering: EFT analysis and nuclear responses}},
  \href{https://doi.org/10.1103/PhysRevD.102.074018}{\emph{Phys. Rev. D}
  {\bfseries 102} (2020) 074018}
  [\href{https://arxiv.org/abs/2007.08529}{{\ttfamily 2007.08529}}].

\bibitem{Miranda:2020tif}
O.G.~Miranda, D.K.~Papoulias, G.~Sanchez~Garcia, O.~Sanders, M.~T\'ortola and
  J.W.F.~Valle, {{Implications of the first detection of coherent elastic
  neutrino-nucleus scattering (CEvNS) with Liquid Argon}},
  \href{https://doi.org/10.1007/JHEP05(2020)130}{\emph{JHEP} {\bfseries 05}
  (2020) 130} [\href{https://arxiv.org/abs/2003.12050}{{\ttfamily
  2003.12050}}].

\bibitem{AtzoriCorona:2022qrf}
M.~Atzori~Corona, M.~Cadeddu, N.~Cargioli, F.~Dordei, C.~Giunti, Y.F.~Li
  et~al., {{Impact of the Dresden-II and COHERENT neutrino scattering data on
  neutrino electromagnetic properties and electroweak physics}},
  \href{https://doi.org/10.1007/JHEP09(2022)164}{\emph{JHEP} {\bfseries 09}
  (2022) 164} [\href{https://arxiv.org/abs/2205.09484}{{\ttfamily
  2205.09484}}].

\bibitem{DeRomeri:2022twg}
V.~De~Romeri, O.G.~Miranda, D.K.~Papoulias, G.~Sanchez~Garcia, M.~T\'ortola and
  J.W.F.~Valle, {{Physics implications of a combined analysis of COHERENT CsI
  and LAr data}}, \href{https://doi.org/10.1007/JHEP04(2023)035}{\emph{JHEP}
  {\bfseries 04} (2023) 035}
  [\href{https://arxiv.org/abs/2211.11905}{{\ttfamily 2211.11905}}].

\bibitem{Breso-Pla:2023tnz}
V.~Bres\'o-Pla, A.~Falkowski, M.~Gonz\'alez-Alonso and K.~Mons\'alvez-Pozo,
  {{EFT analysis of New Physics at COHERENT}},
  \href{https://doi.org/10.1007/JHEP05(2023)074}{\emph{JHEP} {\bfseries 05}
  (2023) 074} [\href{https://arxiv.org/abs/2301.07036}{{\ttfamily
  2301.07036}}].

\bibitem{Breso-Pla:2025cul}
V.~Bres\'o-Pla, S.~Cruz-Alzaga, M.~Gonz\'alez-Alonso and S.~Prakash,
  {{Muon-decay parameters from COHERENT}},
  \href{https://arxiv.org/abs/2502.18175}{{\ttfamily 2502.18175}}.

\bibitem{Chala:2020vqp}
M.~Chala and A.~Titov, {{One-loop matching in the SMEFT extended with a sterile
  neutrino}}, \href{https://doi.org/10.1007/JHEP05(2020)139}{\emph{JHEP}
  {\bfseries 05} (2020) 139}
  [\href{https://arxiv.org/abs/2001.07732}{{\ttfamily 2001.07732}}].

\bibitem{Li:2020lba}
T.~Li, X.-D.~Ma and M.A.~Schmidt, {General neutrino interactions with sterile
  neutrinos in light of coherent neutrino-nucleus scattering and meson
  invisible decays}, \href{https://doi.org/10.1007/JHEP07(2020)152}{\emph{JHEP}
  {\bfseries 07} (2020) 152}
  [\href{https://arxiv.org/abs/2005.01543}{{\ttfamily 2005.01543}}].

\bibitem{ParticleDataGroup:2024cfk}
{\scshape Particle Data Group} collaboration, {{Review of particle physics}},
  \href{https://doi.org/10.1103/PhysRevD.110.030001}{\emph{Phys. Rev. D}
  {\bfseries 110} (2024) }.

\bibitem{Giunti:2024gec}
C.~Giunti, K.~Kouzakov, Y.-F.~Li and A.~Studenikin, {{Neutrino Electromagnetic
  Properties}},  11, 2024.
\newblock 10.1146/annurev-nucl-102122-023242.

\bibitem{Fetscher:1986uj}
W.~Fetscher, H.J.~Gerber and K.F.~Johnson, {{Muon Decay: Complete Determination
  of the Interaction and Comparison with the Standard Model}},
  \href{https://doi.org/10.1016/0370-2693(86)91239-6}{\emph{Phys. Lett. B}
  {\bfseries 173} (1986) 102}.

\bibitem{Fetscher:1994nv}
W.~Fetscher, {{Helicity dependence of the electron-neutrino energy spectrum
  from the decay of unpolarized muons}},
  \href{https://doi.org/10.1103/PhysRevD.49.5945}{\emph{Phys. Rev. D}
  {\bfseries 49} (1994) 5945}.

\bibitem{Pich:2013lsa}
A.~Pich, {{Precision Tau Physics}},
  \href{https://doi.org/10.1016/j.ppnp.2013.11.002}{\emph{Prog. Part. Nucl.
  Phys.} {\bfseries 75} (2014) 41}
  [\href{https://arxiv.org/abs/1310.7922}{{\ttfamily 1310.7922}}].

\bibitem{MuonDecayParameters:2024cfk}
{\scshape Particle Data Group} collaboration, {Muon decay parameters},  in
  \emph{{Review of particle physics}}, vol.~110, pp.~826--828, Phys. Rev. D
  (2024), \href{10.1103/PhysRevD.110.030001}{10.1103/PhysRevD.110.030001}.

\bibitem{Hoferichter:2015dsa}
M.~Hoferichter, J.~Ruiz~de Elvira, B.~Kubis and U.-G.~Mei\ss{}ner,
  {{High-Precision Determination of the Pion-Nucleon \ensuremath{\sigma} Term
  from Roy-Steiner Equations}},
  \href{https://doi.org/10.1103/PhysRevLett.115.092301}{\emph{Phys. Rev. Lett.}
  {\bfseries 115} (2015) 092301}
  [\href{https://arxiv.org/abs/1506.04142}{{\ttfamily 1506.04142}}].

\bibitem{vanKolck:1999mw}
U.~van Kolck, {{Effective field theory of nuclear forces}},
  \href{https://doi.org/10.1016/S0146-6410(99)00097-6}{\emph{Prog. Part. Nucl.
  Phys.} {\bfseries 43} (1999) 337}
  [\href{https://arxiv.org/abs/nucl-th/9902015}{{\ttfamily nucl-th/9902015}}].

\bibitem{Bethe:1936zz}
H.A.~Bethe and R.F.~Bacher, {{Nuclear Physics A. Stationary States of Nuclei}},
  \href{https://doi.org/10.1103/RevModPhys.8.82}{\emph{Rev. Mod. Phys.}
  {\bfseries 8} (1936) 82}.

\bibitem{Falkowski:2021vdg}
A.~Falkowski, M.~Gonz\'alez-Alonso, A.~Palavri\'c and
  A.~Rodr\'\i{}guez-S\'anchez, {{Constraints on subleading interactions in beta
  decay Lagrangian}},
  \href{https://doi.org/10.1007/JHEP02(2024)091}{\emph{JHEP} {\bfseries 02}
  (2024) 091} [\href{https://arxiv.org/abs/2112.07688}{{\ttfamily
  2112.07688}}].

\bibitem{Barranco:2011wx}
J.~Barranco, A.~Bolanos, E.A.~Garces, O.G.~Miranda and T.I.~Rashba, {{Tensorial
  NSI and Unparticle physics in neutrino scattering}},
  \href{https://doi.org/10.1142/S0217751X12501473}{\emph{Int. J. Mod. Phys. A}
  {\bfseries 27} (2012) 1250147}
  [\href{https://arxiv.org/abs/1108.1220}{{\ttfamily 1108.1220}}].

\bibitem{Healey:2013vka}
K.J.~Healey, A.A.~Petrov and D.~Zhuridov, {{Nonstandard neutrino interactions
  and transition magnetic moments}},
  \href{https://doi.org/10.1103/PhysRevD.87.117301}{\emph{Phys. Rev. D}
  {\bfseries 87} (2013) 117301}
  [\href{https://arxiv.org/abs/1305.0584}{{\ttfamily 1305.0584}}].

\bibitem{Papoulias:2015iga}
D.K.~Papoulias and T.S.~Kosmas, {{Neutrino transition magnetic moments within
  the non-standard neutrino\textendash{}nucleus interactions}},
  \href{https://doi.org/10.1016/j.physletb.2015.06.039}{\emph{Phys. Lett. B}
  {\bfseries 747} (2015) 454}
  [\href{https://arxiv.org/abs/1506.05406}{{\ttfamily 1506.05406}}].

\bibitem{Glick-Magid:2023uhk}
A.~Glick-Magid, {{Nonrelativistic nuclear reduction for tensor couplings in
  dark matter direct detection and \ensuremath{\mu}\textrightarrow{}e
  conversion}}, \href{https://doi.org/10.1103/PhysRevD.110.L051701}{\emph{Phys.
  Rev. D} {\bfseries 110} (2024) L051701}
  [\href{https://arxiv.org/abs/2312.08339}{{\ttfamily 2312.08339}}].

\bibitem{Liao:2025hcs}
J.~Liao, J.~Tang and B.-L.~Zhang, {{Tensor interaction in coherent elastic
  neutrino-nucleus scattering}},
  \href{https://arxiv.org/abs/2502.10702}{{\ttfamily 2502.10702}}.

\bibitem{AristizabalSierra:2019zmy}
D.~Aristizabal~Sierra, J.~Liao and D.~Marfatia, {{Impact of form factor
  uncertainties on interpretations of coherent elastic neutrino-nucleus
  scattering data}}, \href{https://doi.org/10.1007/JHEP06(2019)141}{\emph{JHEP}
  {\bfseries 06} (2019) 141}
  [\href{https://arxiv.org/abs/1902.07398}{{\ttfamily 1902.07398}}].

\bibitem{Bischer:2019ttk}
I.~Bischer and W.~Rodejohann, {{General neutrino interactions from an effective
  field theory perspective}},
  \href{https://doi.org/10.1016/j.nuclphysb.2019.114746}{\emph{Nucl. Phys. B}
  {\bfseries 947} (2019) 114746}
  [\href{https://arxiv.org/abs/1905.08699}{{\ttfamily 1905.08699}}].

\bibitem{Papoulias:2019xaw}
D.K.~Papoulias, T.S.~Kosmas and Y.~Kuno, {{Recent probes of standard and
  non-standard neutrino physics with nuclei}},
  \href{https://doi.org/10.3389/fphy.2019.00191}{\emph{Front. in Phys.}
  {\bfseries 7} (2019) 191} [\href{https://arxiv.org/abs/1911.00916}{{\ttfamily
  1911.00916}}].

\bibitem{Chang:2020jwl}
W.-F.~Chang and J.~Liao, {{Constraints on light singlet fermion interactions
  from coherent elastic neutrino-nucleus scattering}},
  \href{https://doi.org/10.1103/PhysRevD.102.075004}{\emph{Phys. Rev. D}
  {\bfseries 102} (2020) 075004}
  [\href{https://arxiv.org/abs/2002.10275}{{\ttfamily 2002.10275}}].

\bibitem{Han:2020pff}
T.~Han, J.~Liao, H.~Liu and D.~Marfatia, {{Scalar and tensor neutrino
  interactions}}, \href{https://doi.org/10.1007/JHEP07(2020)207}{\emph{JHEP}
  {\bfseries 07} (2020) 207}
  [\href{https://arxiv.org/abs/2004.13869}{{\ttfamily 2004.13869}}].

\bibitem{CONUS:2021dwh}
{\scshape CONUS} collaboration, {{Novel constraints on neutrino physics beyond
  the standard model from the CONUS experiment}},
  \href{https://doi.org/10.1007/JHEP05(2022)085}{\emph{JHEP} {\bfseries 05}
  (2022) 085} [\href{https://arxiv.org/abs/2110.02174}{{\ttfamily
  2110.02174}}].

\bibitem{Majumdar:2021vdw}
A.~Majumdar, D.K.~Papoulias and R.~Srivastava, {{Dark matter detectors as a
  novel probe for light new physics}},
  \href{https://doi.org/10.1103/PhysRevD.106.013001}{\emph{Phys. Rev. D}
  {\bfseries 106} (2022) 013001}
  [\href{https://arxiv.org/abs/2112.03309}{{\ttfamily 2112.03309}}].

\bibitem{AristizabalSierra:2022axl}
D.~Aristizabal~Sierra, V.~De~Romeri and D.K.~Papoulias, {{Consequences of the
  Dresden-II reactor data for the weak mixing angle and new physics}},
  \href{https://doi.org/10.1007/JHEP09(2022)076}{\emph{JHEP} {\bfseries 09}
  (2022) 076} [\href{https://arxiv.org/abs/2203.02414}{{\ttfamily
  2203.02414}}].

\bibitem{Chen:2022xkk}
F.-Z.~Chen, M.-D.~Zheng and H.-H.~Zhang, {{Nonperturbative effects in neutrino
  magnetic moments}},
  \href{https://doi.org/10.1103/PhysRevD.106.095009}{\emph{Phys. Rev. D}
  {\bfseries 106} (2022) 095009}
  [\href{https://arxiv.org/abs/2206.13122}{{\ttfamily 2206.13122}}].

\bibitem{Majumdar:2022nby}
A.~Majumdar, D.K.~Papoulias, R.~Srivastava and J.W.F.~Valle, {{Physics
  implications of recent Dresden-II reactor data}},
  \href{https://doi.org/10.1103/PhysRevD.106.093010}{\emph{Phys. Rev. D}
  {\bfseries 106} (2022) 093010}
  [\href{https://arxiv.org/abs/2208.13262}{{\ttfamily 2208.13262}}].

\bibitem{Fridell:2023rtr}
K.~Fridell, L.~Gr\'af, J.~Harz and C.~Hati, {{Probing lepton number violation:
  a comprehensive survey of dimension-7 SMEFT}},
  \href{https://doi.org/10.1007/JHEP05(2024)154}{\emph{JHEP} {\bfseries 05}
  (2024) 154} [\href{https://arxiv.org/abs/2306.08709}{{\ttfamily
  2306.08709}}].

\bibitem{Demirci:2023tui}
M.~Demirci and M.F.~Mustamin, {{Solar neutrino constraints on light mediators
  through coherent elastic neutrino-nucleus scattering}},
  \href{https://doi.org/10.1103/PhysRevD.109.015021}{\emph{Phys. Rev. D}
  {\bfseries 109} (2024) 015021}
  [\href{https://arxiv.org/abs/2312.17502}{{\ttfamily 2312.17502}}].

\bibitem{Chatterjee:2024vkd}
S.S.~Chatterjee, S.~Lavignac, O.G.~Miranda and G.~Sanchez~Garcia, {{Exploring
  the sensitivity to non-standard neutrino interactions of NaI and cryogenic
  CsI detectors at the Spallation Neutron Source}},
  \href{https://doi.org/10.1103/PhysRevD.110.095027}{\emph{Phys. Rev. D}
  {\bfseries 110} (2024) 095027}
  [\href{https://arxiv.org/abs/2402.16953}{{\ttfamily 2402.16953}}].

\bibitem{Karmakar:2024ywn}
S.~Karmakar, M.K.~Singh, S.~Karadaǧ, H.T.~Wong, H.B.~Li, V.~Sharma et~al.,
  {{Search for new physics with reactor neutrino at Kuo-Sheng neutrino
  laboratory}}, \href{https://doi.org/10.1007/s12648-024-03406-x}{\emph{Indian
  J. Phys.} {\bfseries 99} (2025) 1845}.

\bibitem{DeRomeri:2024iaw}
V.~De~Romeri, D.K.~Papoulias and C.A.~Ternes, {{Bounds on new neutrino
  interactions from the first CE$\nu$NS data at direct detection experiments}},
   \href{https://arxiv.org/abs/2411.11749}{{\ttfamily 2411.11749}}.

\bibitem{Chattaraj:2025rtj}
A.~Chattaraj, A.~Majumdar, D.K.~Papoulias and R.~Srivastava, {{Probing
  conventional and new physics at the ESS with coherent elastic
  neutrino-nucleus scattering}},
  \href{https://arxiv.org/abs/2501.12443}{{\ttfamily 2501.12443}}.

\bibitem{Sergio_talk_2024}
\textbf{S. Cruz-Alzaga}, M.~González-Alonso and S.~Prakash, {\textbf{Lightning
  talk and poster:} {EFT} analysis of the {COHERENT} experiment with
  right-handed neutrinos},  in \emph{6th Magnificent CEvNS workshop, Valencia,
  Spain}, 2024-06-12,
  \href{https://indico.cern.ch/event/1342813/contributions/5995195/}{https://indico.cern.ch/event/1342813/contributions/5995195/}.

\bibitem{Lindner:2016wff}
M.~Lindner, W.~Rodejohann and X.-J.~Xu, {{Coherent Neutrino-Nucleus Scattering
  and new Neutrino Interactions}},
  \href{https://doi.org/10.1007/JHEP03(2017)097}{\emph{JHEP} {\bfseries 03}
  (2017) 097} [\href{https://arxiv.org/abs/1612.04150}{{\ttfamily
  1612.04150}}].

\bibitem{Vogel:1989iv}
P.~Vogel and J.~Engel, {{Neutrino Electromagnetic Form-Factors}},
  \href{https://doi.org/10.1103/PhysRevD.39.3378}{\emph{Phys. Rev. D}
  {\bfseries 39} (1989) 3378}.

\bibitem{Tomalak:2020zfh}
O.~Tomalak, P.~Machado, V.~Pandey and R.~Plestid, {{Flavor-dependent radiative
  corrections in coherent elastic neutrino-nucleus scattering}},
  \href{https://doi.org/10.1007/JHEP02(2021)097}{\emph{JHEP} {\bfseries 02}
  (2021) 097} [\href{https://arxiv.org/abs/2011.05960}{{\ttfamily
  2011.05960}}].

\bibitem{Jeong_NuFact2023}
H.~Jeong, ``{CENNS-750, A ton-scale single phase LAr CEvNS detector}.'' August,
  2023.

\bibitem{COHERENT:2022nrm}
{\scshape COHERENT} collaboration, {{The COHERENT Experimental Program}},  in
  \emph{{Snowmass 2021}}, 4, 2022
  [\href{https://arxiv.org/abs/2204.04575}{{\ttfamily 2204.04575}}].

\bibitem{Baxter:2019mcx}
D.~Baxter et~al., {{Coherent Elastic Neutrino-Nucleus Scattering at the
  European Spallation Source}},
  \href{https://doi.org/10.1007/JHEP02(2020)123}{\emph{JHEP} {\bfseries 02}
  (2020) 123} [\href{https://arxiv.org/abs/1911.00762}{{\ttfamily
  1911.00762}}].

\bibitem{Barbeau:2021exu}
P.S.~Barbeau, Y.~Efremenko and K.~Scholberg, {{COHERENT at the Spallation
  Neutron Source}},
  \href{https://doi.org/10.1146/annurev-nucl-101918-023518}{\emph{Ann. Rev.
  Nucl. Part. Sci.} {\bfseries 73} (2023) 41}
  [\href{https://arxiv.org/abs/2111.07033}{{\ttfamily 2111.07033}}].

\bibitem{Abele:2022iml}
H.~Abele et~al., {{Particle Physics at the European Spallation Source}},
  \href{https://doi.org/10.1016/j.physrep.2023.06.001}{\emph{Phys. Rept.}
  {\bfseries 1023} (2023) 1}
  [\href{https://arxiv.org/abs/2211.10396}{{\ttfamily 2211.10396}}].

\bibitem{An:2025lws}
F.~An et~al., {{High-Precision Physics Experiments at Huizhou Large-Scale
  Scientific Facilities}},  \href{https://arxiv.org/abs/2504.21050}{{\ttfamily
  2504.21050}}.

\bibitem{PhysRevLett.81.520}
{\scshape KARMEN} collaboration, {Measurement of the energy spectrum of
  {${\mathit{\ensuremath{\nu}}}_{\mathit{e}}$} from muon decay and implications
  for the lorentz structure of the weak interaction},
  \href{https://doi.org/10.1103/PhysRevLett.81.520}{\emph{Phys. Rev. Lett.}
  {\bfseries 81} (1998) 520}.

\bibitem{MISHRA1990170}
S.~Mishra et~al., {Inverse muon decay, {$\nu_\mu$} + {$e^{-}$} → {$\mu^{-}$}
  + {$\nu_e$}, at the fermilab tevatron},
  \href{https://doi.org/https://doi.org/10.1016/0370-2693(90)91099-W}{\emph{Physics
  Letters B} {\bfseries 252} (1990) 170}.

\bibitem{CHARM-II:1995xfh}
{\scshape CHARM-II} collaboration, {{A Precise measurement of the cross-section
  of the inverse muon decay $\nu_\mu + e^- \rightarrow \mu^- + \nu_e$}},
  \href{https://doi.org/10.1016/0370-2693(95)01298-6}{\emph{Phys. Lett. B}
  {\bfseries 364} (1995) 121}.

\end{thebibliography}\endgroup
}

\end{document}